\documentclass{article}
\usepackage{eurosym}
\usepackage{graphicx}
\usepackage[utf8]{inputenc}
\usepackage[T1]{fontenc}
\usepackage{indentfirst}
\usepackage[margin=0.6in,nomarginpar]{geometry}
\usepackage[final]{hyperref}
\usepackage{amsmath}
\usepackage{hyperref}
\usepackage{cite}
\usepackage{subcaption}
\usepackage{caption}
\usepackage{amssymb}
\usepackage{multirow}

\hypersetup{
colorlinks=true,
linkcolor=blue,
citecolor=blue,
filecolor=magenta,
urlcolor=blue
}

\begin{document}

\title{Dunkl-Pauli Equation in the Presence of a Magnetic Field}
\author{H. Bouguerne\thanks{hacene.bouguerne@univ-oeb.dz} \\
Laboratoire de syst\`{e}mes dynamiques et contr\^{o}le (L.S.D.C), \\
D\'{e}partement des sciences de la mati\`{e}re,\\
Facult\'{e} des Sciences Exactes et SNV,\\
Universit\'{e} de Oum-El-Bouaghi, 04000, Oum El Bouaghi, Algeria. \and B.
Hamil\thanks{%
hamilbilel@gmail.com} \\
Laboratoire de Physique Math\'{e}matique et Subatomique, \\
Facult\'{e}\ des Sciences Exactes, \\
Universit\'{e}\ Constantine 1, Constantine, Algeria. \and B. C. L\"{u}tf\"{u}%
o\u{g}lu \thanks{%
bekir.lutfuoglu@uhk.cz (Corresponding author)} \\
Department of Physics, University of Hradec Kr\'{a}lov\'{e}, \\
Rokitansk\'{e}ho 62, 500 03 Hradec Kr\'{a}lov\'{e}, Czechia. \and M. Merad 
\thanks{%
    meradm@gmail.com} \\
Laboratoire de syst\`{e}mes dynamiques et contr\^{o}le (L.S.D.C), \\
D\'{e}partement des sciences de la mati\`{e}re,\\
Facult\'{e} des Sciences Exactes et SNV,\\
Universit\'{e} de Oum-El-Bouaghi, 04000, Oum El Bouaghi, Algeria.}
\maketitle

\begin{abstract}
The Pauli equation, an important equation of quantum mechanics, allows us to study the dynamics of spin-$1/2$ particles. { The Dunkl derivative, when used instead of the ordinary derivative, leads to obtaining parity-dependent solutions. Motivated by these facts, in this work, we consider a two-dimensional nonrelativistic spin-$1/2$ particle system in the presence of an external magnetic field, and we investigate its parity-dependent dynamics by solving the Pauli equation analytically. Next,} we assume the system to be in thermal equilibrium, {and} we examine various thermal quantities of the system.

\end{abstract}

\section{Introduction}
In literature, particles are classified {according to} various features. One of the most {widely employed}  classifications is based on {particle} spin, which refers to {the} intrinsic angular momentum property. In physics, the terms boson and fermion are used to refer to particles with integer and half-integer {spins}, respectively.  According to quantum statistical mechanics, ensembles that are constituted by bosons or fermions have different energy distribution functions and {therefore different properties.} Strictly speaking, the bosonic particle ensemble obeys Bose-Einstein statistics, while the fermionic particle ensemble accepts Fermi-Dirac statistics. Accordingly, the total wave functions of those ensembles present symmetric or antisymmetric properties with the exchange of identical particles. The Schrödinger equation, {which is} used to determine the dynamics of particles in the non-relativistic regime of quantum mechanics, does not {provide} wave function solutions {that depend on the spin properties of the particles.} To overcome this problem, in 1927 Pauli modified the Schr\"{o}dinger equation by introducing spin matrices, which later became known by his name, {Pauli equation, hereafter} (PE) \cite{Pauli}. A year later, Dirac proposed a first-order differential equation to describe spin$-1/2$ particle dynamics in the relativistic regime \cite{Dirac}. Although the great success of the Dirac equation relegated the PE to the second stage, we observe { its frequent use} in textbooks \cite{0, 1, 2, Berestetskii, 3}. Undoubtedly, the most important reason for this is that the relativistic Dirac equation can be reduced to the PE in the non-relativistic limit. This fact was first proven in 1929 when Breit recommended the omission of small components \cite{Breit}. Then, in 1950{,} Foldy and Wouthuysen showed that such a reduction is also possible under a suitable unitary transformation \cite{FT}. { Almost a century after the publication of the three fundamental equations of quantum mechanics, we see that studies of their correlation are still ongoing} \cite{4, Chapman, 6, 8, 12, 16}. Besides { these efforts, we observe other interesting studies. For example,} Tkachuk, Niederle et al, Ioffe et al and Nikitin used the supersymmetric factorization method to investigate the PE in \cite{17, Niederle99, 15, 10}. Nikitin presented the discrete and Lie symmetry algebras of the PE in  \cite{Nikitin99, Nikitin22}. Karat et al considered an abelian magnetic monopole and derived the self-adjoint extension of the PE \cite{18}. Recently, Cunha et al explored the Aharonov-Bohm effect by employing the self-adjoint extension \cite{Cunha}. Kochan et al handled the PE with complex boundary conditions \cite{11}. Kosugi in \cite{9} and later Wang et al in \cite{7} examined the PE solution for the particles that are confined on curved surfaces. In {another} interesting work, Célérier et al discussed the PE in the framework of fractal geometry \cite{13}. Falek et al presented the solution of the Pauli oscillator in the noncommutative phase space \cite{19}. It is worth noting that in a different context, Dechoum et al presented the derivation of the Pauli equation out of a classical Liouville equation \cite{Dechoum}.

On the other hand, according to the well-known formulation of quantum mechanics, the equations of motion can be determined from commutation relations. Seventy-three years ago Wigner wondered whether the opposite determination might be valid. To understand this, he examined the harmonic oscillator problem and tried to extract commutation relations out of the equations of motion  \cite{Wigner}. He obtained the questioned commutation relationship with an additional constant term and concluded that the inverse determination was possible but not unique. A year later, Yang considered a quantum harmonic oscillator in a suitable tricky Hilbert space description and showed that commutation relations could be derived from the equations of motion at the cost of an additional reflection operator \cite{Yang}. For this purpose, he extended the traditional momentum operator with an extra term containing the reflection operator and a constant called the Wigner parameter. A few decades later, Watanabe demonstrated the self-adjoint extension of Yang's harmonic oscillator Hamiltonian \cite{Watanabe}. For a while, this modified algebra was linked to the quantum Calogero and Calogero-Sutherland-Moser models by relating the Wigner parameter and the Calogero coupling constant \cite{Brzezinski, Plyushchay, Jagannathan}. In 1989, a mathematician, Charles Dunkl, while studying differential-difference and reflection groups and their symmetries, developed Yang's proposal and he defined the differential-difference operator, to be called by his own name, instead of the so-called Yang operator \cite{Dunkl}. {In the following years the} Dunkl derivative has found diverse applications in mathematics \cite{ Heckman, DunklC, Xu} and physics \cite{Pl1, Pl2, Pl3, Pl4, Pl5, Hikami, Kakei, Lapointe, Quesne, Bie, PlyushchayM, Luo, Axel1, Axel2, Axel3, Axel4, Axel5, Axel6, Cancan, new1, quezada}.

We observe that the interest in Dunkl operator-based studies, where the standard derivatives are substituted with Dunkl derivatives,  has increased, especially in the last decade. For example, Genest et al investigated the two-dimensional isotropic and anisotropic Dunkl-harmonic oscillator, respectively in \cite{G1, G2}. Then, they discussed isotropic case symmetry algebra \cite{G3} and presented its three-dimensional solution \cite{G4}. Salazar-Ramirez et al revisited the same problem in 2017 and investigated their coherent states in two dimensions \cite{Ramirez1}. Besides these non-relativistic regime oscillators, Dirac, Klein-Gordon, and Duffin-Kemmer-Petiau oscillators were also examined in the Dunkl formalism. For instance, the Dunkl-Dirac oscillator solutions were discussed widely in \cite{Sargol, Mota1, Bilel3, Ojeda, BilalEPL}, while the Dunkl-Klein-Gordon and  Dunkl-Duffin-Kemmer-Petiau oscillator solutions were presented in \cite{Mota1, Bilel3, BilalEPL, Mota2, Mota3, Bilel1} and \cite{Merad}, respectively. The Coulomb and Mie-type interactions were investigated within the Dunkl formalism in \cite{Genest2015Coulomb, Ramirez2, Ghaz, Ghaz1, Kim, Ghaz2, MotGul, MotMie}. Moreover, the Dunkl-electrostatics and Maxwell equations were formed \cite{Jang, ChungDunkl}. In other interesting works, the fundamentals of Dunkl-Newtonian, Dunkl-quantum, and Dunkl-statistical mechanics were explored  \cite{Chungrev, Chungqm, Dong, Marcelo, superstat}. A graphene layer's thermal quantities {were} given in the Dunkl formalism in \cite{Bilel2}. Dunkl-Bose-Einstein condensation of an ideal Bose gas was explored for various trapped potential energies in \cite{BEC1, BEC2, BEC3, BEC4}. Very recently, the Dunkl derivative was employed to investigate the dynamics of relativistic particles in curved space \cite{DRelat1} and thermal quantities of black holes \cite{DRelat2}.

To our best knowledge, the PE has not been investigated in the context of Dunkl formalism. To fill this gap, in this paper, we consider a two-dimensional system of spin-$1/2$ particles, which are under the effect of an external magnetic field {to} examine their non-relativistic dynamics with the Dunkl-Pauli equation (DPE). {Moreover, we} aim to derive several thermal quantities and compare {our findings} with the ordinary case. {To this end, we} construct the manuscript as follows: In the following section, we establish the DPE in two dimensions and explore the solution set that consists of eigenvalue and eigenfunctions depending on the reflection and Wigner parameters. In section 3, we derive several thermal quantities within the Dunkl formalism, namely the Helmholtz free energy, mean energy, specific heat, and entropy functions, by constructing the partition function. Finally, we briefly conclude the manuscript {in} the conclusion section.

\section{Dunkl-Pauli equation in two dimensions}

{We start with the stationary form of the PE, which is the non-relativistic limit of the Dirac equation in standard quantum mechanics, }
\begin{eqnarray}
\frac{1}{2m}\left( \overrightarrow{\mathbf{\pi }} 
 \cdot \overrightarrow{\sigma }%
\right) ^{2}\psi =E\psi .
\end{eqnarray}
Here, { the conjugate momentum stands as:}
\begin{eqnarray}
    \pi _{j}=p_{j}-\frac{e}{c}A_{j}, \quad j=1,2,
\end{eqnarray}
{ with the magnetic vector potential, $A_{j}$, the Pauli matrices, $\sigma _{j}$, and the spinor wave function with two components}
\begin{eqnarray}
  \psi =\left( 
\begin{array}{c}
\Phi \\ 
\Psi%
\end{array}
\right).   
\end{eqnarray}
{To} obtain the DPE, we replace the ordinary momentum operator with the Dunkl momentum operator 
\begin{equation}
p_{j}=\frac{1}{i}D_{j},
\end{equation}
where the Dunkl derivatives
\begin{eqnarray}
D_{j}=\frac{\partial }{\partial x_{j}}+\frac{\nu _{j}}{x_{j}}\left(
1-R_{j}\right),  \label{D}
\end{eqnarray}
with the reflection operators
\begin{eqnarray}
 R_{j}=\left( -1\right) ^{x_{j}\partial
_{x_{j}}}, \text{ \ \ }{ \partial _{x_{j}}=\frac{\partial }{\partial x_{j}} },  \label{Dr}
\end{eqnarray}
substitute the ordinary partial derivatives. {Here we should emphasize that the Wigner parameters should carry out the condition  $\nu _{j}>-1/2$,} \cite{Chungqm}. In the Dunkl formalism, the reflection operators satisfy the following actions
\begin{equation}
R_{j}f\left( x_{j}\right) =f\left( -x_{j}\right) ;\text{ \ \ }%
R_{j}R_{i}=R_{i}R_{j}\text{\ };\text{ \ \ }R_{j}x_{i}=-\delta
_{ij}x_{i}R_{j}, \label{R}
\end{equation}
and the deformed Heisenberg algebra obeys
\begin{equation}
\left[ x_{i},D_{j}\right] =\delta _{ij}\left( 1+2\nu _{j}R_{j}\right) ;\text{
\ \ }\left[ D_{i},D_{j}\right] =0;\text{ \ \ }\left[ x_{i},x_{j}\right] =0.
\label{C}
\end{equation}
With the latter definitions, the
Dunkl-Pauli Hamiltonian takes the form 
\begin{equation}
H=\frac{1}{2m}\left( \overrightarrow{\mathbf{\pi }} 
 \cdot \overrightarrow{\sigma }%
\right) ^{2}=\frac{1}{2m}\mathbf{\pi }_{1}^{2}+\frac{1}{2m}\mathbf{\pi }%
_{2}^{2}+\frac{1}{2m}\left[ \mathbf{\pi }_{1};\mathbf{\pi }_{2}\right]
\sigma _{1}\sigma _{2}.
\end{equation}
Here, we choose the vector potential in the symmetric gauge
\begin{equation}
\overrightarrow{A}=\frac{B}{2}\left( -x_{2}\,{\hat{i}}+x_{1}\,
{\hat{j}}\right) ,
\end{equation}%
with a magnetic field of magnitude $B$, { and ${(\,\hat{i},\,\hat{j}\,)}$ are Cartesian orthonormal basis}. Then, by using the Dunkl-deformed algebra we obtain the Dunkl-Pauli Hamiltonian terms as follows:
\begin{eqnarray}
\frac{\mathbf{\pi }_{1}^{2}+\mathbf{\pi }_{2}^{2}}{2m}&=&-\frac{1}{2m}\bigg(\Delta _{D}-\frac{B^{2}e^{2}}{4c^{2}}\left( x_{1}^{2}+x_{2}^{2}\right) +\frac{Be}{ic}\left(x_{1}D_{2}-x_{2}D_{2}\right)\bigg), \\
\frac{1}{2m}\left[ \mathbf{\pi }_{1};\mathbf{\pi }_{2}\right] &=&-\frac{Be}{2imc}\bigg(1+\nu _{1}R_{1}+\nu _{2}R_{2}\bigg) ,
\end{eqnarray}%
with the Dunkl-Laplacian operator
\begin{equation}
\Delta _{D}=\frac{\partial ^{2}}{\partial x_{1}^{2}}+\frac{\partial ^{2}}{\partial x_{2}{}^{2}}+\frac{2\nu _{1}}{x_{1}}\frac{\partial }{\partial x_{1}}+\frac{2\nu _{2}}{x_{2}}\frac{\partial }{\partial x_{2}}-\frac{\nu _{1}}{x_{1}^{2}}\left( 1-R_{1}\right) -\frac{\nu _{2}}{x_{2}^{2}}\left(1-R_{2}\right).
\end{equation}
{Then}, we organize these terms and arrive at the final form of the Dunkl-Pauli Hamiltonian  
\begin{equation}
H=-\frac{1}{2m}\Delta _{D}+\frac{m\omega _{c}^{2}}{8}\left(
x_{1}^{2}+x_{2}^{2}\right) +i\frac{\omega _{c}}{2}\left(
x_{1}D_{2}-x_{2}D_{2}\right) -\mu _{B}\left( 1+\nu _{1}R_{1}+\nu
_{2}R_{2}\right) {\mathbf{\overrightarrow{B} \cdot \overrightarrow{\sigma}} ,}
\end{equation}
in terms of the Bohr magneton,  $\mu _{B}=\frac{\left\vert e\right\vert }{2mc}$, and isotropic harmonic oscillator frequency, $\omega _{c}=\frac{Be}{mc}$. We observe that for $\nu_1=\nu_2=0$, the Dunkl-Pauli Hamiltonian reduces to the ordinary Pauli Hamiltonian \cite{19}.
{ Hereafter we decide to use polar coordinates, $( r , \theta)$, for convenience since the system is two-dimensional  }
\begin{equation}
x_{1}=r\cos \theta, \text{ \ and } x_{2}=r\sin \theta .
\end{equation}%
In this manner, the Dunkl-Pauli Hamiltonian reads:
\begin{equation}
H=-\frac{1}{2m}\frac{\partial ^{2}}{\partial r^{2}}-\frac{1+2\nu
_{1}+2\nu _{2}}{2mr}\frac{\partial }{\partial r}+\frac{m\omega _{c}^{2}}{8}%
r^{2}+\frac{\mathcal{B}_{\theta }}{mr^{2}}+\frac{\omega _{c}}{2}\mathcal{J}%
_{\theta }-g_{s}\mu _{B}\left( 1+\nu _{1}R_{1}+\nu _{2}R_{2}\right) \mathbf{%
B\cdot S,}
\end{equation}%
where $\mathbf{S=}\frac{\mathbf{\sigma }}{2}$ is the spin-1/2 operator, $g_{s}$ is the free electron $g-$factor ($g_{s}=2.0023$). Here, $\mathcal{B}_{\theta }$ and $\mathcal{J}_{\theta }$ denote the Dunkl angular operators that are defined by 
\begin{eqnarray}
\mathcal{B}_{\theta } &=&-\frac{1}{2}\frac{\partial ^{2}}{\partial \theta
^{2}}+\left( \nu _{1}\tan \theta -\nu _{2}\cot \theta \right) \frac{\partial 
}{\partial \theta }+\frac{\nu _{1}}{2\cos ^{2}\theta }\left( 1-R_{1}\right) +%
\frac{\nu _{2}}{2\sin ^{2}\theta }\left( 1-R_{2}\right) , \\
\mathcal{J}_{\theta } &=&i\left( \frac{\partial }{\partial \theta }+\nu
_{2}\cot \theta \left( 1-R_{2}\right) -\nu _{1}\tan \theta \left(
1-R_{1}\right) \right) .
\end{eqnarray}
In the latter coordinate system, {we take} the wave function in the form of
\begin{equation}
\psi _{m_{s}}=\phi _{m_{s}}\left( r,\theta \right) \chi _{m_{s}},  \label{f}
\end{equation}
where $\chi _{m_{s}}$ corresponds to the spin function,%
\begin{equation}
\mathbf{S}_{z}\chi _{m_{s}}=\frac{m_{s}}{2}\chi _{m_{s}};\text{ \ }m_{s}=\pm
1;\text{ \ \ }\chi _{+1}=\left( 
\begin{array}{c}
1 \\ 
0%
\end{array}%
\right) ;\text{ \ \ }\chi _{-1}=\left( 
\begin{array}{c}
0 \\ 
1%
\end{array}%
\right) .  \label{s}
\end{equation}
It is worth mentioning that in the polar coordinate system, the reflection operators act on a function, $\phi \left( r,\theta\right) $,  as follows:
\begin{equation}
R_{1}\phi \left( r,\theta \right) =\phi \left( r,\pi -\theta \right) ;\text{
\ \ \ }R_{2}\phi \left( r,\theta \right) =\phi \left( r,-\theta \right) .
\end{equation}%
Next, we employ the relation  
\begin{equation}
\mathcal{J}_{\theta }^{2}=2\mathcal{B}_{\theta }+2\nu _{1}\nu _{2}\left(1-R_{1}R_{2}\right) ,  \label{a}
\end{equation}%
and substitute  { Eqs.} (\ref{f}), (\ref{s}) and (\ref{a}),
into the Dunkl-Pauli equation. We obtain
\small
\begin{eqnarray}
\left[ \frac{\partial ^{2}}{\partial r^{2}}+\frac{1+2\nu _{1}+2\nu _{2}}{r}%
\frac{\partial }{\partial r}-\frac{m^{2}\omega _{c}^{2}r^{2}}{4}-\frac{%
\mathcal{J}_{\theta }^{2}-2\nu _{1}\nu _{2}\left( 1-R_{1}R_{2}\right) }{r^{2}%
}-m\omega _{c}\mathcal{J}_{\theta }+m B \mu _{B} g_{s} m_{s} \left( 1+\nu_{1}R_{1}+\nu _{2}R_{2}\right) +2mE\right] \phi _{m_{s}}=0\mathbf{.}
\label{ET}
\end{eqnarray}%
\normalsize
In order to derive a solution, first we have to address the eigenvalues and eigenfunctions of the Dunkl-angular operator $\mathcal{J}_{\theta }$, as it has been done in the previous papers \cite{G1, Bilel2}. Here, we closely follow \cite{G1} and summarize the solution's key points. Since the operator $R_{1}R_{2}$  commutes with the operator $\mathcal{J}_{\theta }$, we consider a separable wave function of the form
\begin{equation}
\phi _{\epsilon ,m_{s}}\left( r,\theta \right) =\digamma _{m_{s}}\left(
r\right) \Theta _{\epsilon }\left( \theta \right) ,
\end{equation}
where $\lambda _{\epsilon }$ is the eigenvalue of Dunkl-angular operator 
\begin{equation}
\mathcal{J}_{\theta }\Theta _{\epsilon }\left( \theta \right) =\lambda
_{\epsilon }\Theta _{\epsilon }\left( \theta \right) .  \label{E}
\end{equation}
Here, $\epsilon =\epsilon _{1}\epsilon _{2}=\pm 1$, and $\epsilon _{1}$, $\epsilon _{2}$ are the eigenvalues of the reflection operators $R_{1}$ and $R_{2}$, respectively. {Next, we} determine the eigenfunctions, $\Theta _{\epsilon }\left( \theta \right) $, and the eigenvalues, $\lambda _{\epsilon }$, within two different cases:

\paragraph{First case $\epsilon =+1:$}

This case is characterized by $\epsilon _{1}=\epsilon _{2}=+1$ or $\epsilon
_{1}=\epsilon _{2}=-1$, and the solutions of Eq. (\ref{E}) are given by
\begin{equation}
\Theta _{+1}\left( \theta \right) =a_{\ell }\mathbf{P}_{\ell }^{\left( \nu
_{1}+1/2,\nu _{2}+1/2\right) }\left( -2\cos \theta \right) \pm a_{\ell
}^{^{\prime }}\sin \theta \cos \theta \, \mathbf{P}_{\ell -1}^{\left( \nu
_{1}+1/2,\nu _{2}+1/2\right) }\left( -2\cos \theta \right) ,
\end{equation}%
and 
\begin{equation}
\lambda _{+}=\pm 2\sqrt{\ell \left( \ell +\nu _{1}+\nu _{2}\right) },\text{
\ \ \ \ \ \ }\ell \in \mathbf{N}^{\ast }.
\end{equation}%
Here, $\mathbf{P}_{\ell -1}^{\left( \alpha ,\beta \right) }\left( x\right) $
are the Jacobi polynomials. If we assume that $\digamma _{m_{s}}\left(
r\right) =\digamma _{m_{s}}^{\epsilon _{1},\epsilon _{2}}\left( r\right) $, then
Eq. (\ref{ET}) turns to%
\begin{equation}
\left[ \frac{\partial ^{2}}{\partial r^{2}}+\frac{1+2\nu _{1}+2\nu _{2}}{r}%
\frac{\partial }{\partial r}-\frac{m^{2}\omega _{c}^{2}r^{2}}{4}-\frac{%
\lambda _{+}^{2}}{r^{2}}-m\omega _{c}\lambda _{+}+mB\mu _{B}g_{s}m_{s}\left(
1+\nu _{1}\epsilon _{1}+\nu _{2}\epsilon _{2}\right) +2mE\right] \digamma
_{\ell ,m_{s}}^{\epsilon _{1},\epsilon _{2}}\left( r\right) =0\mathbf{.}
\label{pE}
\end{equation}%
We observe that there are two possibilities, namely $\epsilon _{1}=\epsilon
_{2}=+1$ and $\epsilon _{1}=\epsilon _{2}=-1$. So, the solution of the above
differential equation splits into two-parity sectors which can be labeled by
the eigenvalues of the quantum numbers, $\epsilon _{1}$ and $\epsilon _{2}$.

\begin{description}
\item[\textbf{1.}] For $\epsilon _{1}=\epsilon _{2}=+1,$ we have%
\begin{equation}
\left[ \frac{\partial ^{2}}{\partial r^{2}}+\frac{1+2\nu _{1}+2\nu _{2}}{r}%
\frac{\partial }{\partial r}-\frac{m^{2}\omega _{c}^{2}r^{2}}{4}-\frac{%
\lambda _{+}^{2}}{r^{2}}-m\omega _{c}\lambda _{+}+mB\mu _{B}g_{s}m_{s}\left(
1+\nu _{1}+\nu _{2}\right) +2mE\right] \digamma _{\ell ,m_{s}}^{+,+}\left(
r\right) =0\mathbf{,}  \label{RE}
\end{equation}%
{and the} solution to Eq. (\ref{pE}) reads:
{
\begin{align}
\digamma_{n,\ell,m_{s}}^{+,+}  & =\mathcal{C}_{n,\ell,m_{s}}^{+,+}%
e^{-\frac{m\omega_{c}}{4}r^{2}}r^{2\ell}\times\nonumber\\
& _{1}\mathbf{F}_{1}\left(  \frac{1+\lambda_{+}+\sqrt{\left(  \nu_{1}+\nu
_{2}\right)  ^{2}+\lambda_{+}^{2}}}{2}-\frac{m_{s}\left(  1+\nu_{1}+\nu
_{2}\right)  }{2}-\frac{E}{\omega_{c}},1+\sqrt{\left(  \nu_{1}+\nu_{2}\right)
^{2}+\lambda_{+}^{2}};\frac{m\omega_{c}}{2}r^{2}\right)  ,
\end{align}
}
where { $_{1}\mathbf{F}_{1}\left( a,b,x\right) $} is the confluent hypergeometric
function, and $\mathcal{C}_{n,\ell ,m_{s}}^{+,+}$ is the normalization constant. For $a=-n$ (non-positive integer), the confluent hypergeometric function reduces to the Laguerre polynomial. Here, the Dunkl-energy eigenvalue function appears as 
{
\begin{equation}
E_{n,\ell ,m_{s}}^{+,+}=\omega _{c}\left( n+\frac{1+\lambda _{+}+\nu
_{1}+\nu _{2}+2\ell }{2}-\frac{m_{s}\left( 1+\nu _{1}+\nu _{2}\right) }{2}%
\right) .  \label{E1}
\end{equation}}
\end{description}

\begin{description}
\item[\textbf{2.}] For $\epsilon _{1}=\epsilon _{2}=-1,$ {we get}
\begin{equation}
\left[ \frac{\partial ^{2}}{\partial r^{2}}+\frac{1+2\nu _{1}+2\nu _{2}}{r}%
\frac{\partial }{\partial r}-\frac{m^{2}\omega _{c}^{2}r^{2}}{4}-\frac{%
\lambda _{+}^{2}}{r^{2}}-m\omega _{c}\lambda _{+}+mB\mu _{B}g_{s}m_{s}\left(
1-\nu _{1}-\nu _{2}\right) +2mE\right] \digamma _{\ell ,m_{s}}^{-,-}\left(
r\right) =0\mathbf{.}  \label{rff}
\end{equation}%
In this case, the radial spinor stands for
{
\begin{align}
\digamma_{n,\ell,m_{s}}^{-,-}\left(  r\right)    & =\mathcal{C}_{n,\ell,m_{s}%
}^{-,-}e^{-\frac{m\omega_{c}}{4}r^{2}}r^{2\ell}\times\nonumber\\
& _{1}\mathbf{F}_{1}\left(  \frac{1+\lambda_{+}+\sqrt{\left(  \nu_{1}+\nu
_{2}\right)  ^{2}+\lambda_{+}^{2}}}{2}-\frac{m_{s}\left(  1-\nu_{1}-\nu
_{2}\right)  }{2}-\frac{E}{\omega_{c}},1+\sqrt{\left(  \nu_{1}+\nu_{2}\right)
^{2}+\lambda_{+}^{2}};\frac{m\omega_{c}}{2}r^{2}\right)  .
\end{align}
}
with the respective energy spectrum function 
{
\begin{equation}
E_{n,\ell ,m_{s}}^{-,-}=\omega _{c}\left( n+\frac{1+\nu _{1}+\nu _{2}+2\ell
+\lambda _{+}}{2}-\frac{m_{s}\left( 1-\nu _{1}-\nu _{2}\right) }{2}\right) .  \label{E2}
\end{equation}}
\end{description}

\paragraph{Second case $\epsilon =-1$}

This case can be examined in two spare cases, namely $\left( \epsilon _{1};\epsilon _{2}\right) =\left( +1,-1\right) $ or $\left( \epsilon _{1};\epsilon _{2}\right) =\left( -1,+1\right)$. {Here, the} angular eigenfunction $\Theta _{-1}\left( \theta \right) $ reads:
{
\begin{equation}
\Theta_{-1}\left(  \theta\right)  =b_{\ell}\cos\theta \,\mathbf{P}_{\ell
-1/2}^{\left(  \nu_{1}+1/2,\nu_{2}-1/2\right)  }\left(  -2\cos\theta\right)
\pm b_{\ell}^{^{\prime}}\sin\theta \, \mathbf{P}_{\ell-1/2}^{\left(  \nu
_{1}-1/2,\nu_{2}+1/2\right)  }\left(  -2\cos\theta\right)  ,
\end{equation}
}

with%
{
\begin{equation}
\lambda_{-}=\pm2\sqrt{\left(  \ell+\nu_{1}\right)  \left(  \ell+\nu
_{2}\right)  },\text{ \ \ \ \ }\ell\in\left\{  1/2,3/2,5/2,...\right\}  .
\end{equation}
}

Under this consideration, Eq. (\ref{ET}) turns to

\begin{equation}
\left[ \frac{\partial ^{2}}{\partial r^{2}}+\frac{1+2\nu _{1}+2\nu _{2}}{r}%
\frac{\partial }{\partial r}-\frac{m^{2}\omega _{c}^{2}r^{2}}{4}-\frac{%
\lambda _{-}^{2}-4\nu _{1}\nu _{2}}{r^{2}}-m\omega _{c}\lambda _{-}+mB\mu
_{B}g_{s}m_{s}\left( 1+\nu _{1}\epsilon _{1}+\nu _{2}\epsilon _{2}\right)
+2mE\right] \digamma _{\ell ,m_{s}}^{\epsilon _{1},\epsilon _{2}}\left(
r\right) =0\mathbf{.}  \label{Epm}
\end{equation}

\begin{description}
\item[\textbf{1.}] For $\epsilon _{1}=+1$ and $\epsilon _{2}=-1$, the
functions $\digamma _{\ell ,m_{s}}^{+,-}\left( r\right) $ explicitly reads:
{
\begin{align}
\digamma_{n,\ell,m_{s}}^{+,-}\left(  r\right)    & =\mathcal{C}_{n,\ell,m_{s}%
}^{+1,-1}e^{-\frac{m\omega_{c}}{4}r^{2}}r^{2\ell} \times \nonumber\\
& _{1}\mathbf{F}_{1}\left(  \frac{1+\lambda_{-}+\sqrt{\left(  \nu_{1}-\nu
_{2}\right)  ^{2}+\lambda_{-}^{2}}}{2}-\frac{m_{s}\left(  1+\nu_{1}-\nu
_{2}\right)  }{2}-\frac{E}{\omega_{c}},1+\sqrt{\left(  \nu_{1}-\nu_{2}\right)
^{2}+\lambda_{-}^{2}};\frac{m\omega_{c}}{2}r^{2}\right)  ,
\end{align}%
}
and the energy spectrum stands for
{
\begin{equation}
E_{n,\ell ,m_{s}}^{+,-}=\omega _{c}\left( n+\frac{1+\lambda _{-}+\nu
_{1}+\nu _{2}+2\ell }{2}-\frac{m_{s}\left( 1+\nu _{1}-\nu _{2}\right) }{2}%
\right) .  \label{E3}
\end{equation}}

\item[\textbf{2.}] For $\epsilon _{1}=-1$ and $\epsilon _{2}=+1$, the eigensolution $\digamma _{\ell ,m_{s}}^{-,+}\left( r\right) $ becomes
{
\begin{align}
\digamma_{n,\ell,m_{s}}^{-,+}\left(  r\right)    & =\mathcal{C}_{n,\ell,m_{s}%
}^{-1,+1}e^{-\frac{m\omega_{c}}{4}r^{2}}r^{2\ell}  \times 
 \nonumber\\
& _{1}\mathbf{F}_{1}\left(  \frac{1+\lambda_{-}+\sqrt{\left(  \nu_{2}-\nu
_{1}\right)  ^{2}+\lambda_{-}^{2}}}{2}-\frac{m_{s}\left(  1+\nu_{2}-\nu
_{1}\right)  }{2}-\frac{E}{\omega_{c}},1+\sqrt{\left(  \nu_{2}-\nu_{1}\right)
^{2}+\lambda_{-}^{2}};\frac{m\omega_{c}}{2}r^{2}\right)  ,
\end{align}%
}
and the Dunkl-energy eigenvalue function appears
{
\begin{equation}
E_{n,\ell ,m_{s}}^{-,+}\left( r\right) =\omega _{c}\left( n+\frac{1+\lambda
_{-}+\nu _{1}+\nu _{2}+2\ell }{2}-\frac{m_{s}\left( 1+\nu _{2}-\nu
_{1}\right) }{2}\right).  \label{E4}
\end{equation}}
\end{description}
We notice that the energy levels of the examined system are influenced by
the Wigner parameters $\nu _{j}$. This is a natural consequence of the
modified Heisenberg algebra.

\section{Thermal properties}

\begin{figure}[htb]
\begin{minipage}[t]{0.5\textwidth}
        \centering
        \includegraphics[width=\textwidth]{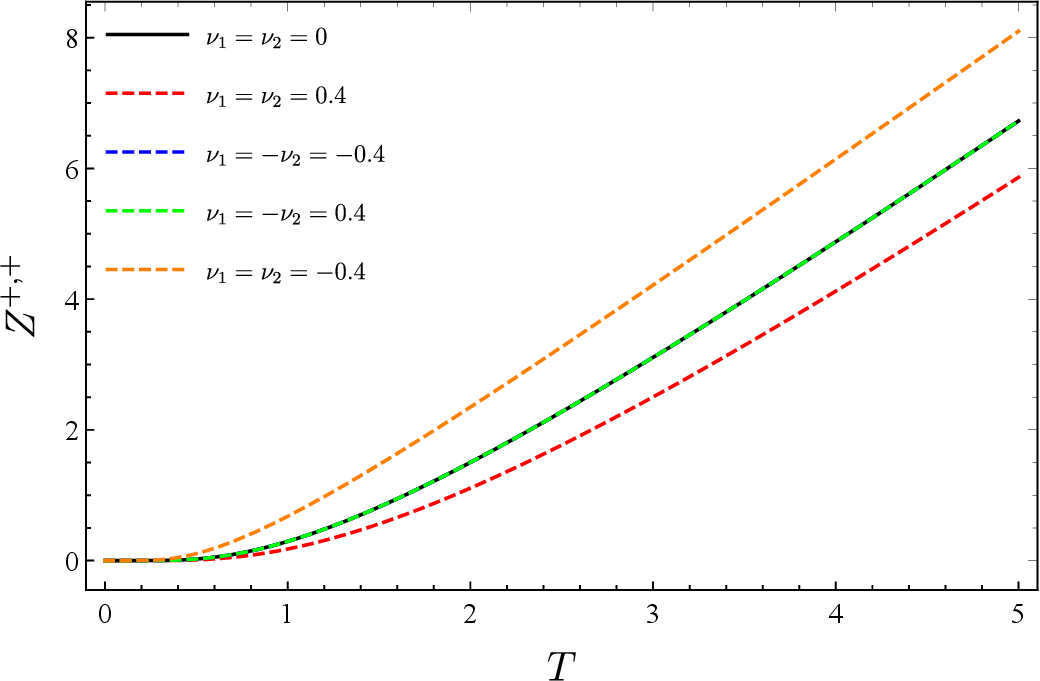}
        \subcaption{$\epsilon _{1}=\epsilon
_{2}=+1$ case. }
           \end{minipage}%
\begin{minipage}[t]{0.50\textwidth}
        \centering
       \includegraphics[width=\textwidth]{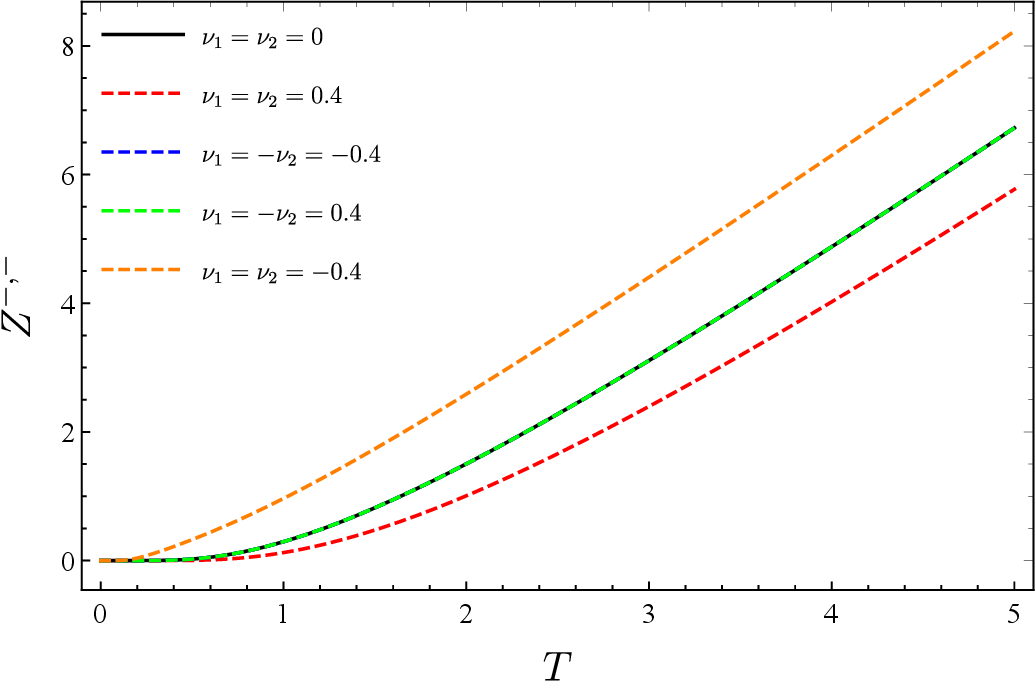}
          \subcaption{ $\epsilon _{1}=\epsilon
_{2}=-1$ case.}
    \end{minipage}\hfill 
\begin{minipage}[t]{0.5\textwidth}
        \centering
        \includegraphics[width=\textwidth]{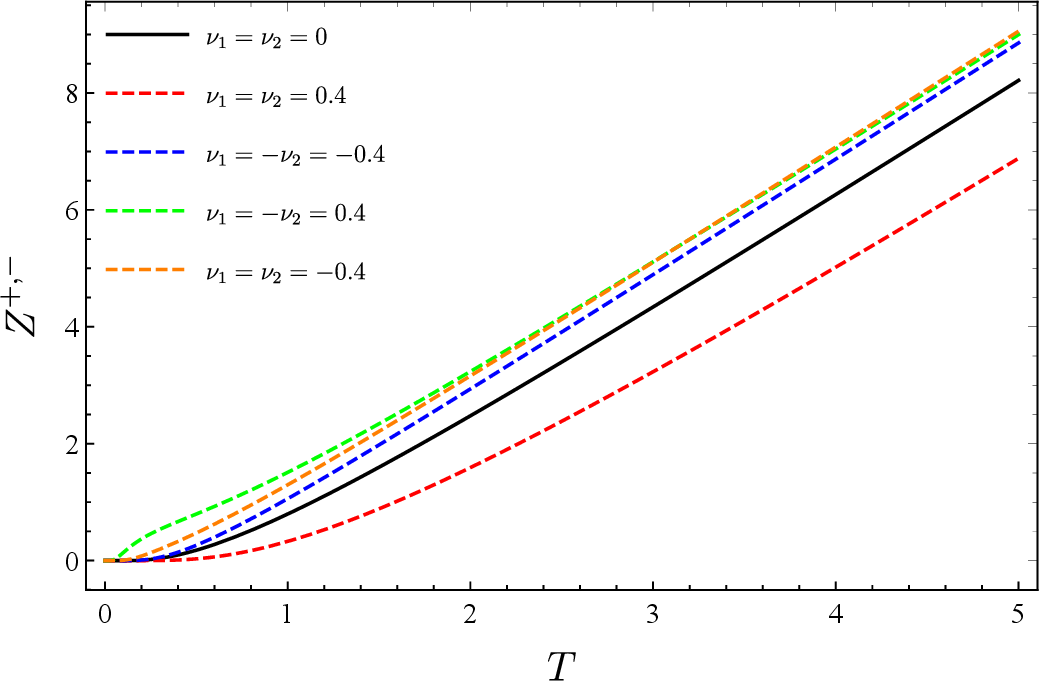}
   \subcaption{ $\epsilon _{1}=+1$, $\epsilon
_{2}=-1$ case.}
   \end{minipage}%
\begin{minipage}[t]{0.50\textwidth}
        \centering
       \includegraphics[width=\textwidth]{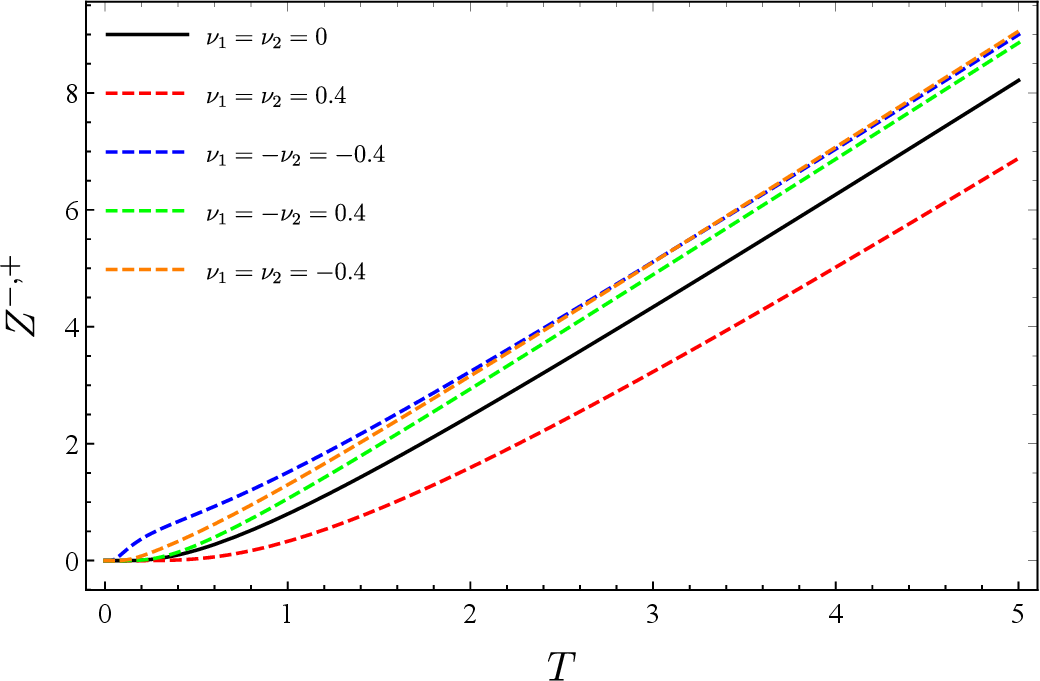}
          \subcaption{$\epsilon _{1}=-1$, $\epsilon_{2}=+1$ case. }
    \end{minipage}\hfill
\caption{A qualitative representation of the Dunkl-Partition function versus temperature for different values of 
$\protect\nu _{1}$ and $\protect\nu _{2}$.} \label{Fig1}
\end{figure}

\begin{figure}[htb]
\begin{minipage}[t]{0.5\textwidth}
        \centering
        \includegraphics[width=\textwidth]{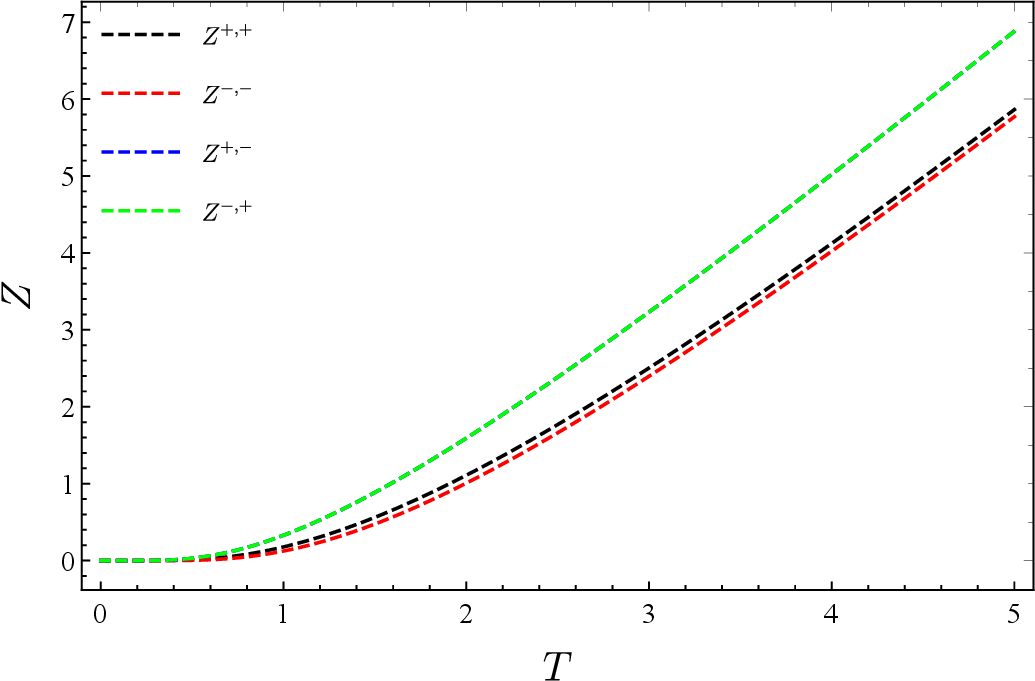}
       \subcaption{ $ \nu_{1}=0.4$, and $\nu_{2}=0.4$.}\label{fig:Ma}
   \end{minipage}%
\begin{minipage}[t]{0.50\textwidth}
        \centering
       \includegraphics[width=\textwidth]{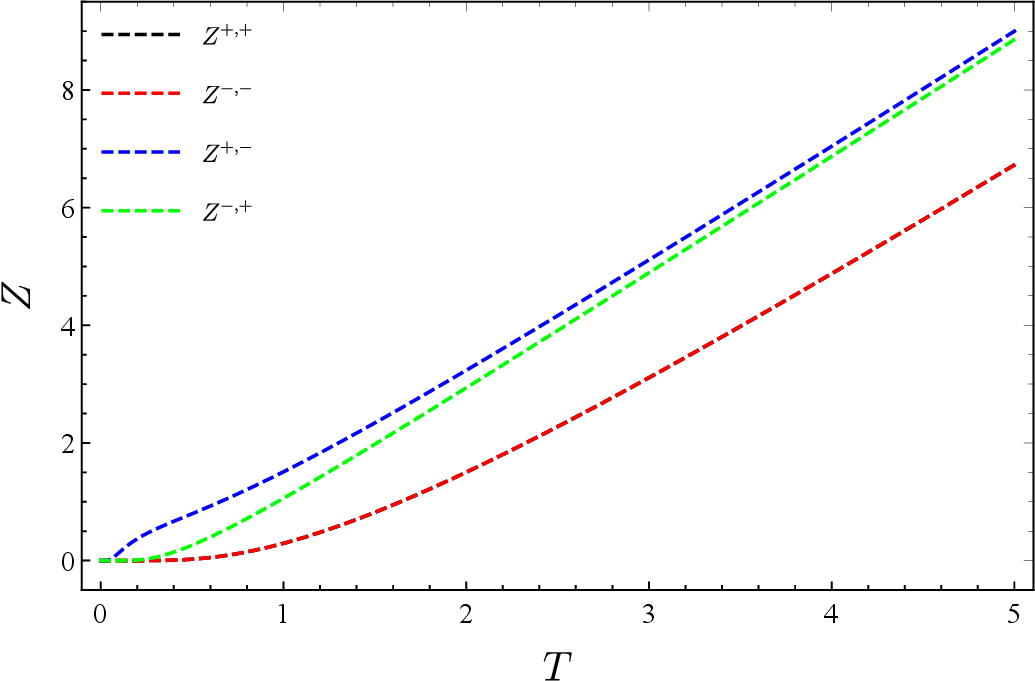}\\
        \subcaption{  $ \nu_{1}=0.4$, and $\nu_{2}=-0.4$.}\label{fig:Mb}
    \end{minipage}\hfill 
\begin{minipage}[b]{0.5\textwidth}
        \centering
        \includegraphics[width=\textwidth]{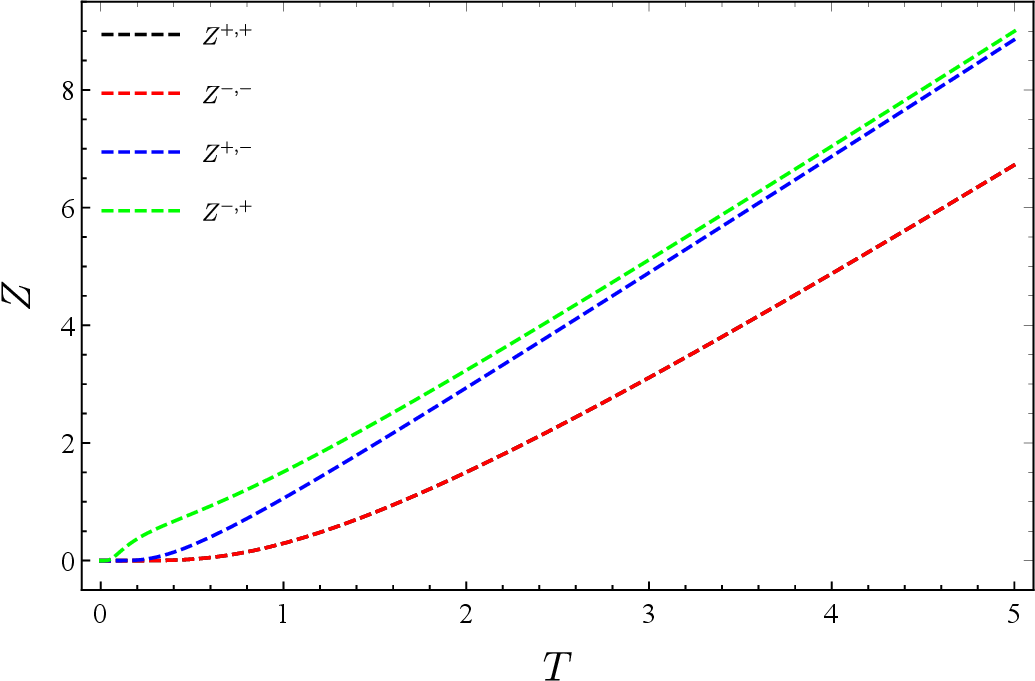}
       \subcaption{ $ \nu_{1}=-0.4$, and $\nu_{2}=0.4$.}\label{fig:Mc}
   \end{minipage}%
\begin{minipage}[b]{0.50\textwidth}
        \centering
       \includegraphics[width=\textwidth]{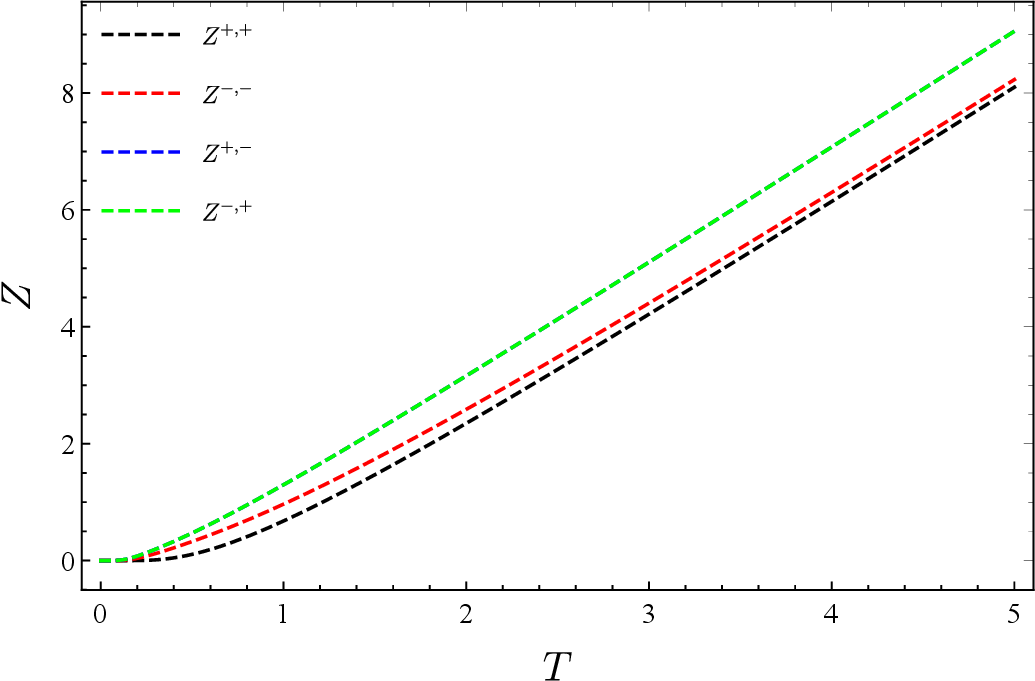}\\
        \subcaption{ $ \nu_{1}=-0.4$, and $\nu_{2}=-0.4$.}\label{fig:Md}
    \end{minipage}\hfill
\caption{A qualitative representation of the Dunkl-partition function versus temperature for constant Wigner parameters.}  \label{Fig2}
\end{figure}

{It is a very well-known} fact that an electron gas adheres to the Fermi-Dirac quantum statistics. Nevertheless, in high-temperature regimes or when dealing with an electron gas at low density, the Maxwell-Boltzmann statistics can be employed as an alternative. In this section, we intend to present the thermal quantities of a system that is described with the DPE. To initiate our calculations, we compute the partition function, $Z$, of an individual particle for {a fixed value of angular momentum,} $\ell$, { \cite{rn1, rn2, rn3, rn4, rn5, rn6, rn7, rn8, rn9}.}
\begin{equation}
Z=\sum_{n=0}^{+\infty }\sum_{m_{s=-1}}^{1}e^{-\beta E_{n,m_{s}}}.
\end{equation}%
Here, $\beta =1/KT,$ { where} $K$ denotes the Boltzmann constant, $T$ represents the thermodynamic temperature and $E_{n,m_{s}}$ {corresponds to} the energy eigenvalues function. { For convenience, we express the obtained energy eigenvalue function, given in Eqs. (\ref{E1}), (\ref{E2}), (\ref{E3}) and (\ref{E4}),} as follows:%
\begin{equation}
E_{n,\ell ,m_{s}}^{\epsilon _{1},\epsilon _{2}}=\omega _{c}\left( n+\frac{1}{%
2}\right) +\omega _{c}\rho _{\ell }^{\epsilon }-\omega _{c}m_{s}\eta
^{\epsilon _{1},\epsilon _{2}},
\end{equation}%
where%
\begin{equation}
\left\{ 
\begin{array}{c}
\rho _{\ell }^{+1}=\frac{\lambda _{+1}+\sqrt{\left( \nu _{1}+\nu _{2}\right)
^{2}+\lambda _{+1}^{2}}}{2} \\ 
\rho _{\ell }^{-1}=\frac{\lambda _{-1}+\sqrt{\left( \nu _{1}-\nu _{2}\right)
^{2}+\lambda _{-1}^{2}}}{2}%
\end{array}%
\right. ,
\end{equation}%
and%
\begin{equation}
\left\{ 
\begin{array}{c}
\eta ^{+1,+1}=\frac{1+\nu _{1}+\nu _{2}}{2} \\ 
\eta ^{-1,-1}=\frac{1-\nu _{1}-\nu _{2}}{2} \\ 
\eta ^{+1,-1}=\frac{1+\nu _{1}-\nu _{2}}{2} \\ 
\eta ^{-1,+1}=\frac{1-\nu _{1}+\nu _{2}}{2} 
\end{array}%
\right. {.}
\end{equation}%
{Then, we} rewrite the partition function in the following form 
\begin{equation}
Z_{\ell }^{\epsilon _{1},\epsilon _{2}}=e^{-\beta \omega _{c}\rho _{\ell
}^{\epsilon }}\sum_{n=0}^{+\infty }e^{-\beta \omega _{c}\left( n+\frac{1}{2}%
\right) }\sum_{m_{s=-1}}^{1}e^{\beta \omega _{c}m_{s}\eta ^{\epsilon
_{1},\epsilon _{2}}}.
\end{equation}%
For {the} calculation, we {employ} the geometric series summation rule
\begin{equation}
\sum_{n=0}^{+\infty }x^{n}=\frac{1}{1-x},
\end{equation}%
so that, we {obtain} the partition function in the subsequent manner
\begin{equation}
Z_{\ell }^{\epsilon _{1},\epsilon _{2}}=e^{-\beta \omega _{c}\rho _{\ell
}^{\epsilon }}\frac{\cosh \left( \beta \omega _{c}\eta ^{\epsilon
_{1},\epsilon _{2}}\right) }{\sinh \left( \frac{\beta \omega _{c}}{2}\right) 
}.  \label{pa}
\end{equation}
In Figs. \ref{Fig1} and \ref{Fig2},  we depict the partition function versus temperature with different Wigner parameter values.

\newpage

\newpage
We observe that the partition function for parity systems with $\epsilon =\pm 1$ increases monotonically with temperature. Furthermore, at a fixed temperature, the partition function decreases as $\nu_{j}$ increases. Additionally, we notice differences in the partition function for parity systems with $\epsilon =\pm 1$ at relatively high temperatures. Moreover, we observe that for systems with even parity ($\epsilon =+ 1$) and $\nu_{1} = -\nu_{2}$, the Dunkl partition function aligns with the ordinary partition function.

\begin{figure}[htb]
\begin{minipage}[t]{0.5\textwidth}
        \centering
        \includegraphics[width=\textwidth]{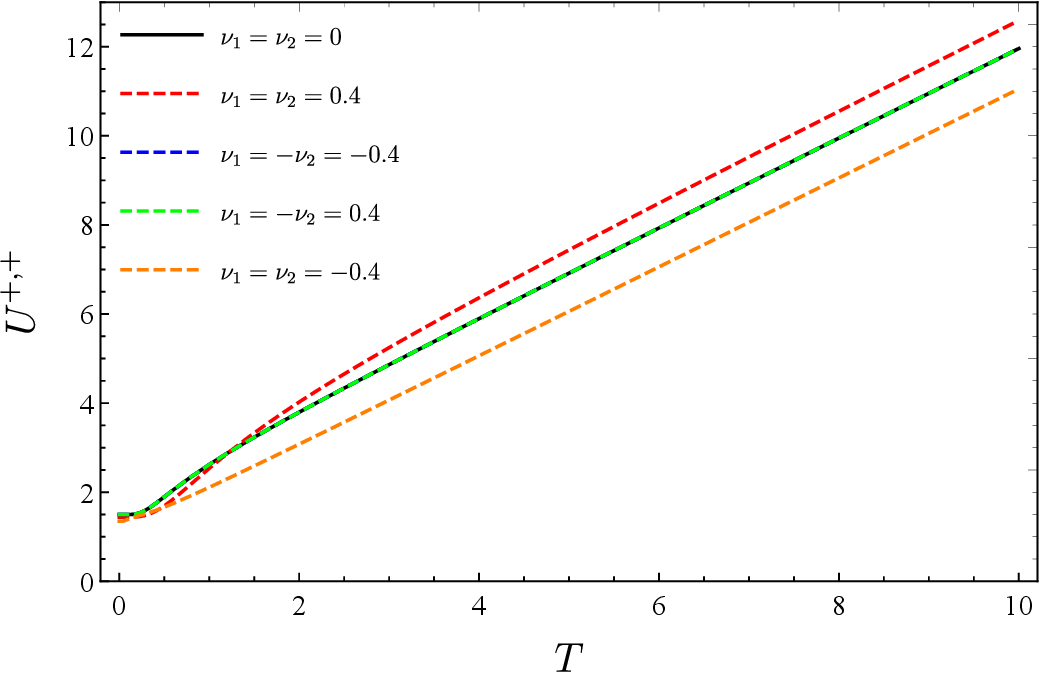}
        \subcaption{$\epsilon _{1}=\epsilon
_{2}=+1$ case. }
           \end{minipage}%
\begin{minipage}[t]{0.50\textwidth}
        \centering
       \includegraphics[width=\textwidth]{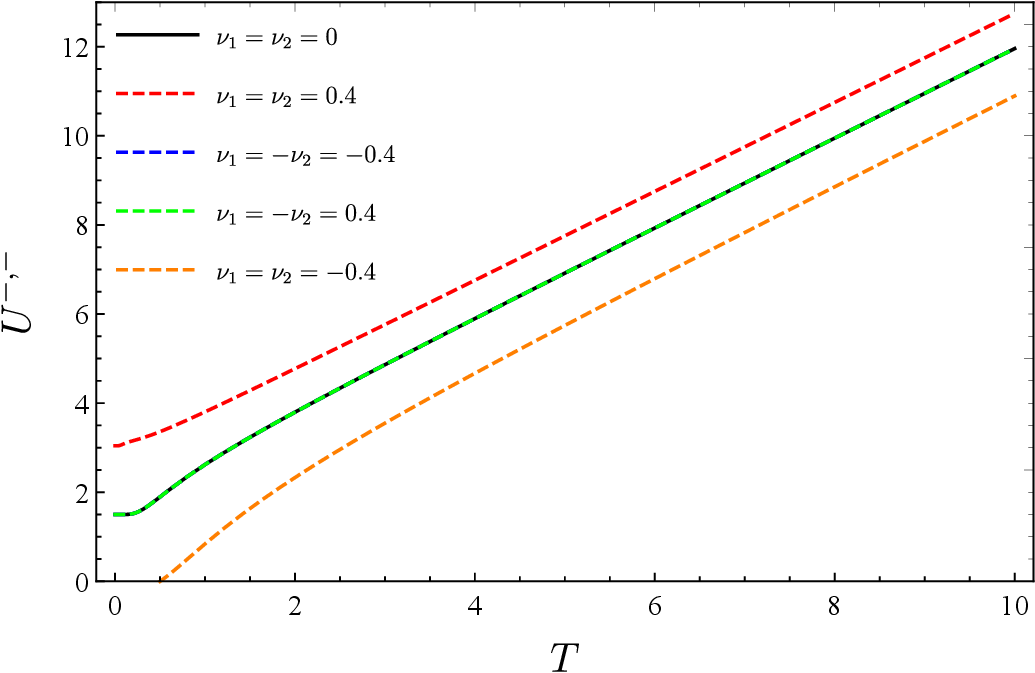}
       \subcaption{$\epsilon _{1}=\epsilon
_{2}=-1$ case. }
    \end{minipage}\hfill 
\begin{minipage}[t]{0.5\textwidth}
        \centering
        \includegraphics[width=\textwidth]{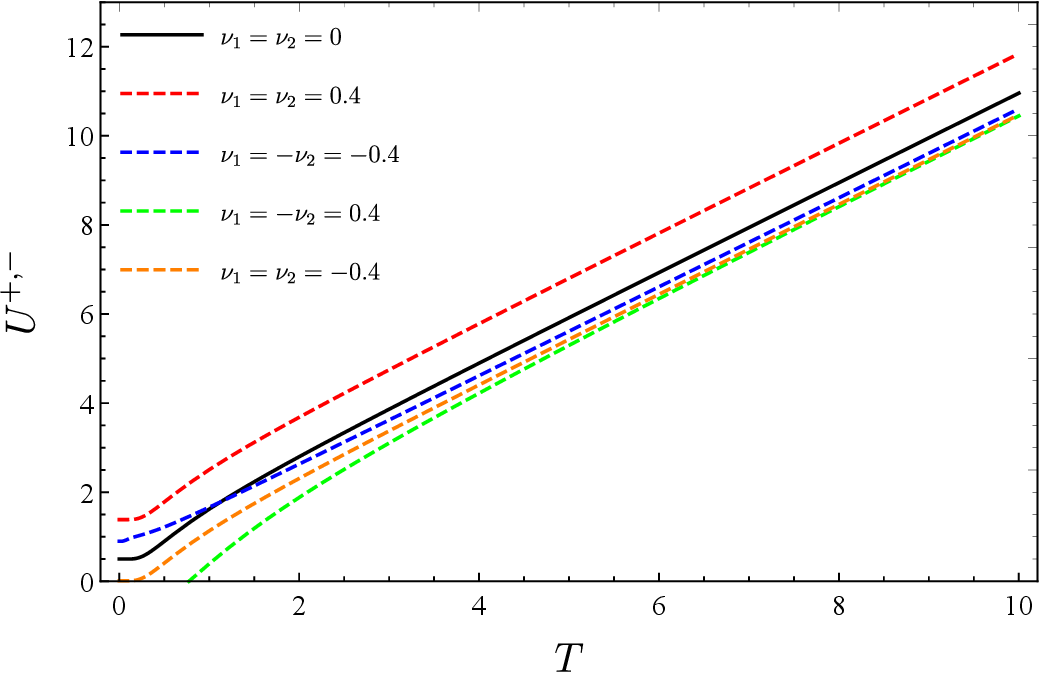}
\subcaption{ $\epsilon _{1}=+1$, $\epsilon
_{2}=-1$ case.}
   \end{minipage}%
\begin{minipage}[t]{0.50\textwidth}
        \centering
       \includegraphics[width=\textwidth]{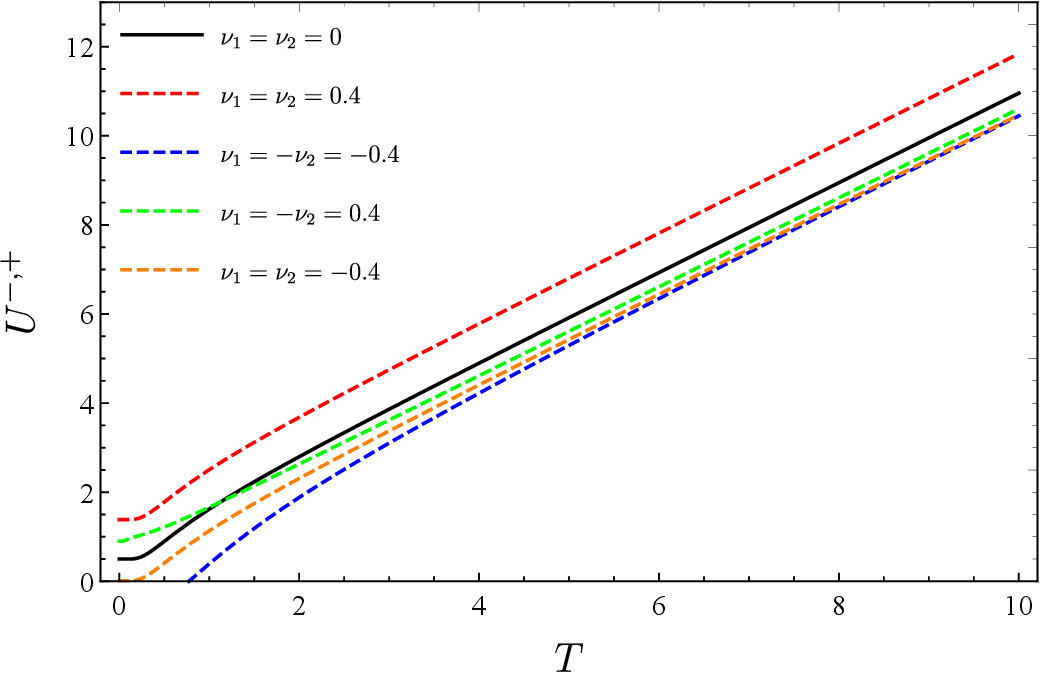}
       \subcaption{ $\epsilon _{1}=-1$, $\epsilon
_{2}=+1$ case.}
    \end{minipage}\hfill
\caption{A qualitative representation of the Dunkl-internal energy versus temperature for different values of $%
\protect\nu _{1}$ and $\protect\nu _{2}$.} \label{Fig3}
\end{figure}

\begin{figure}[htb]
\begin{minipage}[t]{0.5\textwidth}
        \centering
        \includegraphics[width=\textwidth]{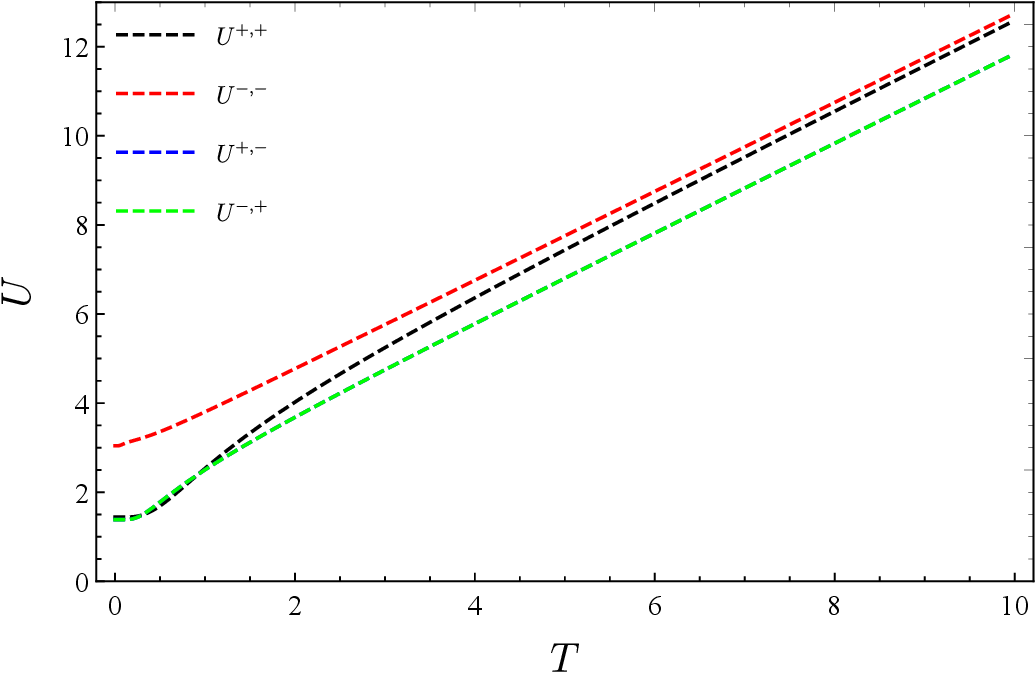}
       \subcaption{ $ \nu_{1}=0.4$, and $\nu_{2}=0.4$.}\label{fig:Ua}
   \end{minipage}%
\begin{minipage}[t]{0.50\textwidth}
        \centering
       \includegraphics[width=\textwidth]{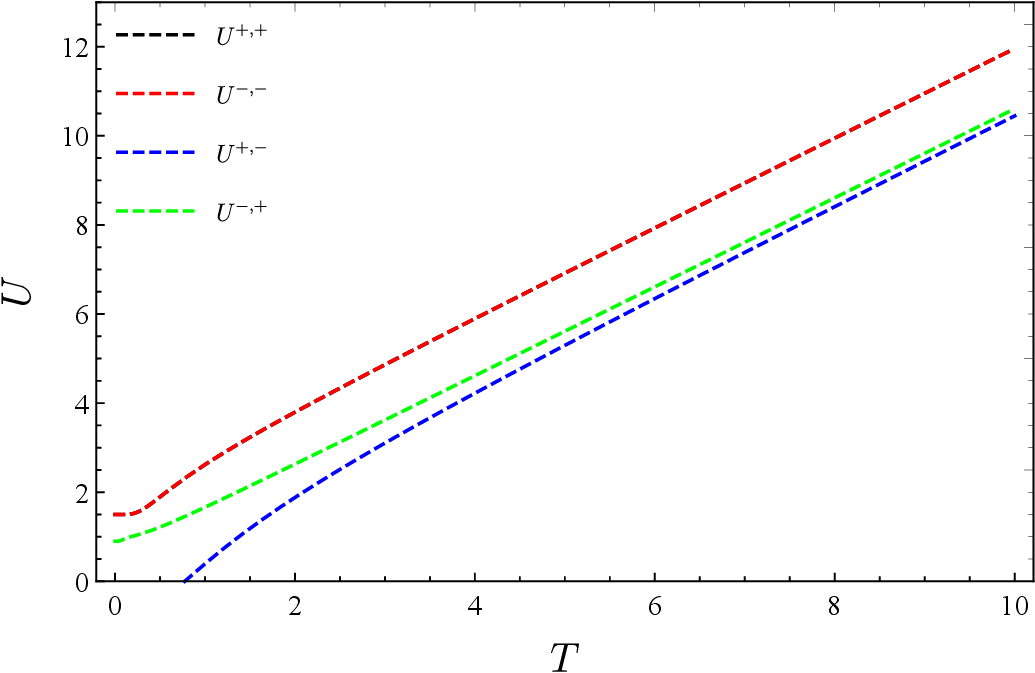}\\
        \subcaption{  $ \nu_{1}=0.4$, and $\nu_{2}=-0.4$.}\label{fig:Ub}
    \end{minipage}\hfill 
\begin{minipage}[b]{0.5\textwidth}
        \centering
        \includegraphics[width=\textwidth]{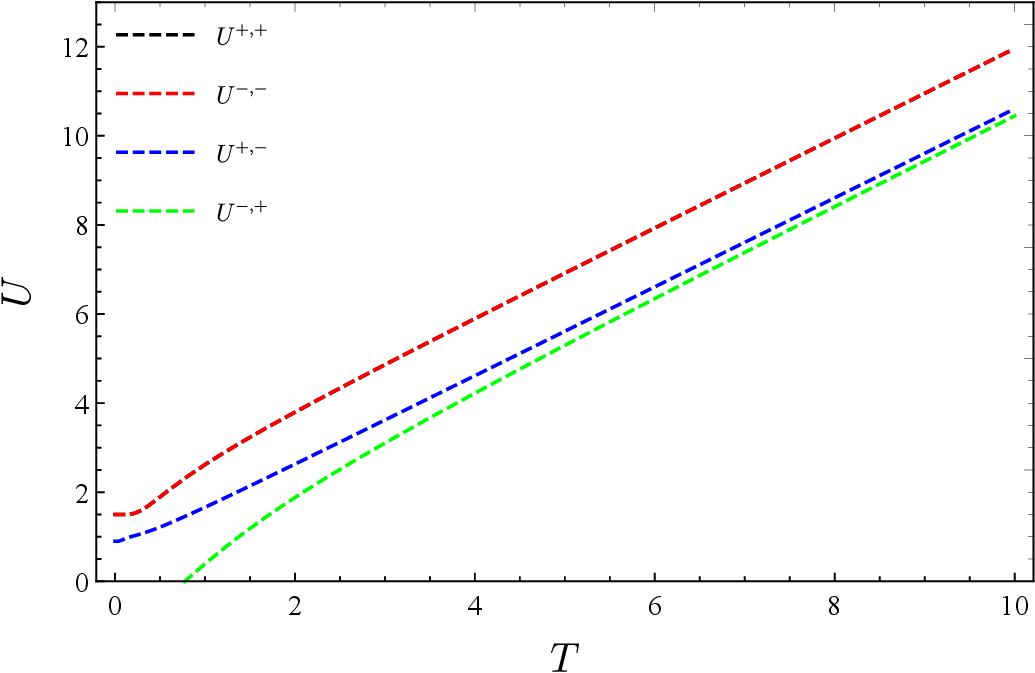}
       \subcaption{ $ \nu_{1}=-0.4$, and $\nu_{2}=0.4$.}\label{fig:Uc}
   \end{minipage}%
\begin{minipage}[b]{0.50\textwidth}
        \centering
       \includegraphics[width=\textwidth]{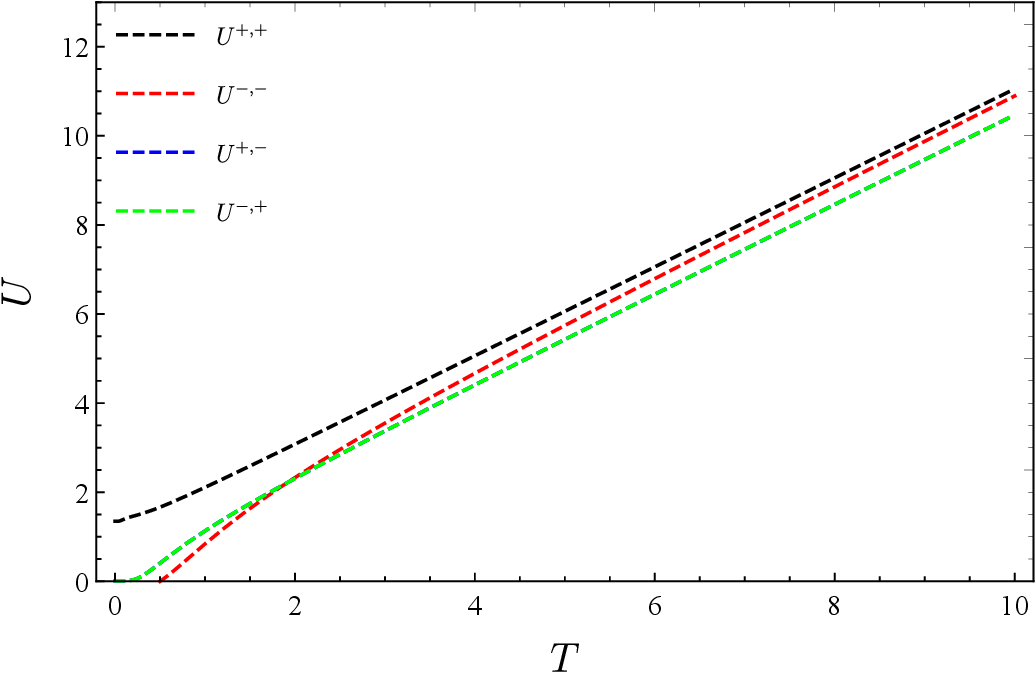}\\
        \subcaption{ $ \nu_{1}=-0.4$, and $\nu_{2}=-0.4$.}\label{fig:Ud}
    \end{minipage}\hfill
\caption{A qualitative representation of the Dunkl-internal energy function versus temperature for constant Wigner parameters. } \label{Fig4}
\end{figure}

\newpage
Then, utilizing the partition function, we proceed to calculate the thermal functions, beginning with the computation of the Dunkl-Helmholtz free
energy function via the following definition:
\begin{equation}
F=-\frac{1}{\beta }\log Z.  \label{DF}
\end{equation}%
By substituting Eq. (\ref{pa}) into Eq. (\ref{DF}), we find%
\begin{equation}
F_{\ell }^{\epsilon _{1},\epsilon _{2}}=\frac{1}{\beta }\log \sinh \left( 
\frac{\beta \omega _{c}}{2}\right) -\frac{1}{\beta }\log \cosh \left( \beta
\omega _{c}\eta ^{\epsilon _{1},\epsilon _{2}}\right) +\omega _{c}\rho
_{\ell }^{\epsilon }. \label{hfe}
\end{equation}%
Here, the first term represents conventional Helmholtz free energy, while the second and third terms show the modification of the Dunkl formalism. Next, by employing 
\begin{eqnarray}
  U=-\frac{\partial }{\partial \beta }\log Z , 
\end{eqnarray}
we derive the Dunkl-internal energy 
\begin{equation}
U_{\ell }^{\epsilon _{1},\epsilon _{2}}=\frac{\omega _{c}}{2}\coth \left( 
\frac{\beta \omega _{c}}{2}\right) -\eta ^{\epsilon _{1},\epsilon
_{2}}\omega _{c}\tanh \left( \beta \omega _{c}\eta ^{\epsilon _{1},\epsilon
_{2}}\right) -\omega _{c}\rho _{\ell }^{\epsilon }.  \label{DU}
\end{equation}%
{Here, we employ the following formulas for discussing the high and low-temperature characteristics:}
\begin{equation}
\tanh x\simeq \left\{ 
\begin{array}{c}
x\text{ \ \ for }x<<1 \\ 
1\text{ \ \ for }x>>1%
\end{array}%
\right. ; \quad \text{ }\coth x\simeq \left\{ 
\begin{array}{c}
\frac{1}{x}\text{ \ \ for }x<<1 \\ 
1\text{ \ \ for }x>>1%
\end{array}%
\right. ,
\end{equation}
and we get
\begin{equation}
U_{\ell }^{\epsilon _{1},\epsilon _{2}}\simeq \left\{ 
\begin{array}{ll}
KT-\omega _{c}\rho _{\ell }^{\epsilon }, & KT>>\omega _{c} \\ 
\left( \frac{1}{2}-\eta ^{\epsilon _{1},\epsilon _{2}}-\rho _{\ell
}^{\epsilon }\right) \omega _{c},  & KT<<\omega _{c}%
\end{array}%
\right. 
\end{equation}%
We observe that in the low-temperature regime, almost all oscillators are in their ground states, while at the high-temperature regime, the first term denotes the usual internal energy, and the second represents the Dunkl modification term. In Figs. \ref{Fig3} and \ref{Fig4},  we show how the Dunkl-internal energy function varies versus temperature with different Wigner parameter values.

{In} Fig. \ref{Fig3}, we observe that the Dunkl-internal energy increases monotonically with temperature in all cases. We also see that for a fixed value of {temperature}, the Dunkl-internal energy increases when the deformation parameter $\nu_{1}+\nu_{2}$ grows. 

\begin{figure}[htb]
\begin{minipage}[t]{0.5\textwidth}
        \centering
        \includegraphics[width=\textwidth]{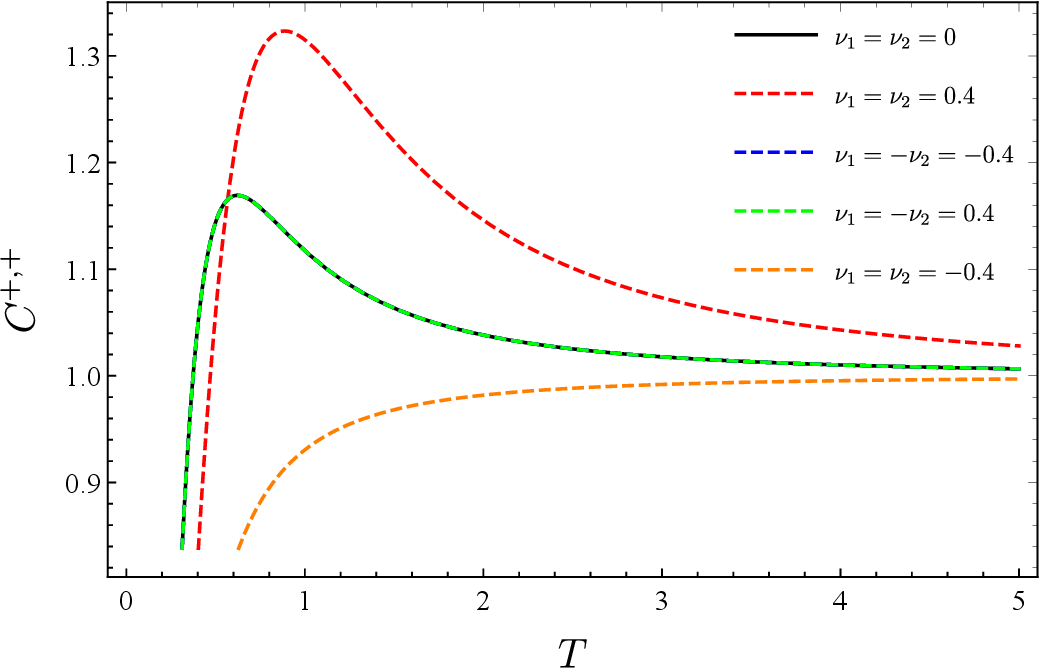}
         \subcaption{$\epsilon _{1}=\epsilon
_{2}=+1$ case. }
           \end{minipage}%
\begin{minipage}[t]{0.50\textwidth}
        \centering
       \includegraphics[width=\textwidth]{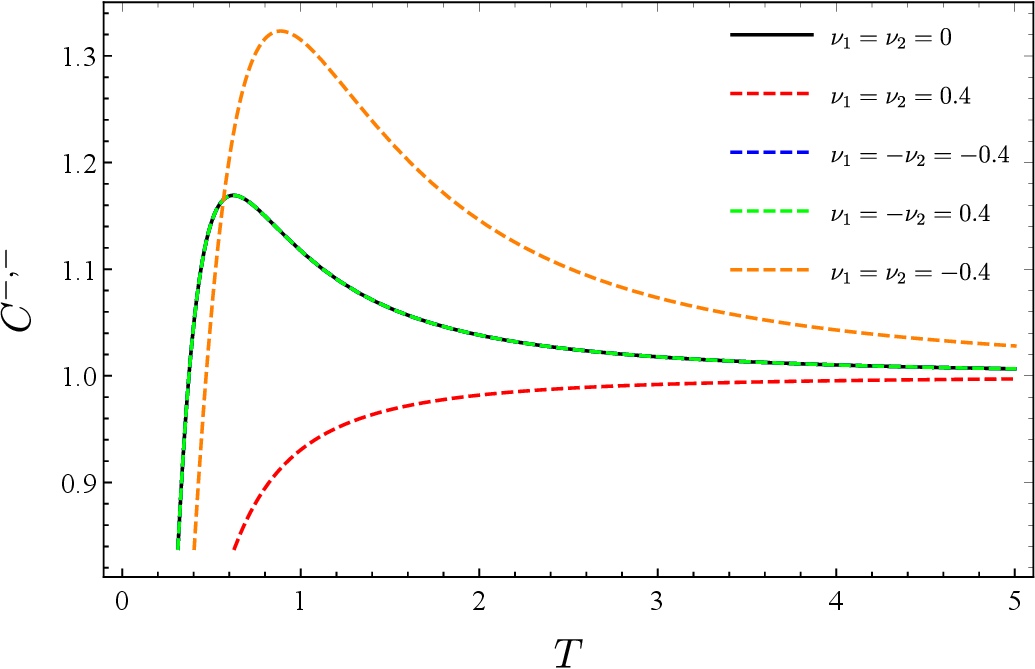}
        \subcaption{$\epsilon _{1}=\epsilon
_{2}=-1$ case. }
    \end{minipage}\hfill 
\begin{minipage}[t]{0.5\textwidth}
        \centering
        \includegraphics[width=\textwidth]{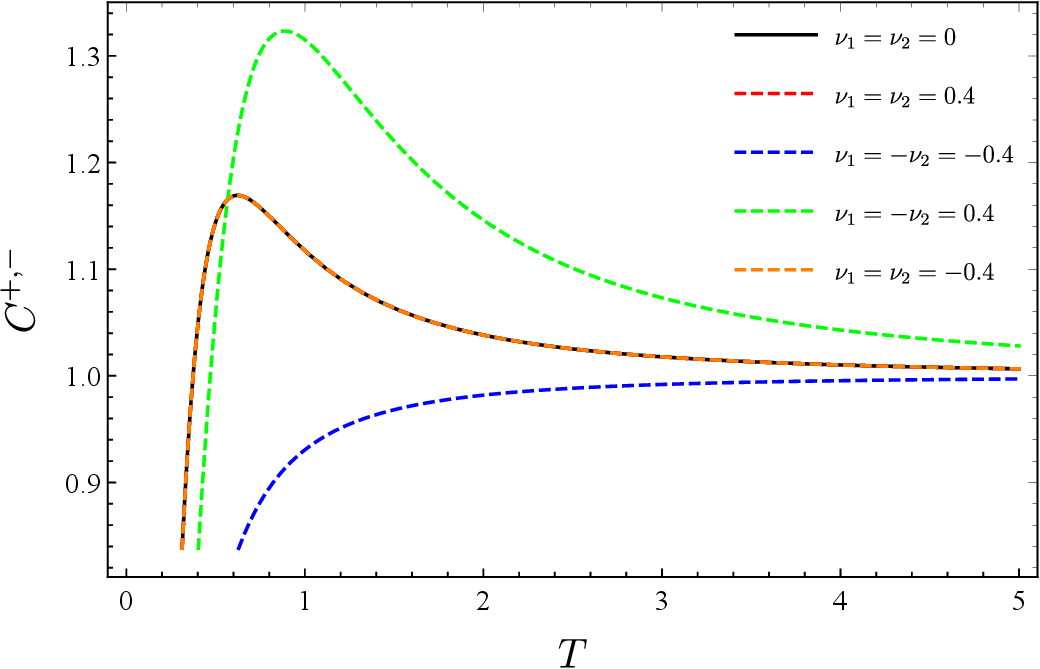}
\subcaption{ $\epsilon _{1}=+1$, $\epsilon
_{2}=-1$ case.}
   \end{minipage}%
\begin{minipage}[t]{0.50\textwidth}
        \centering
       \includegraphics[width=\textwidth]{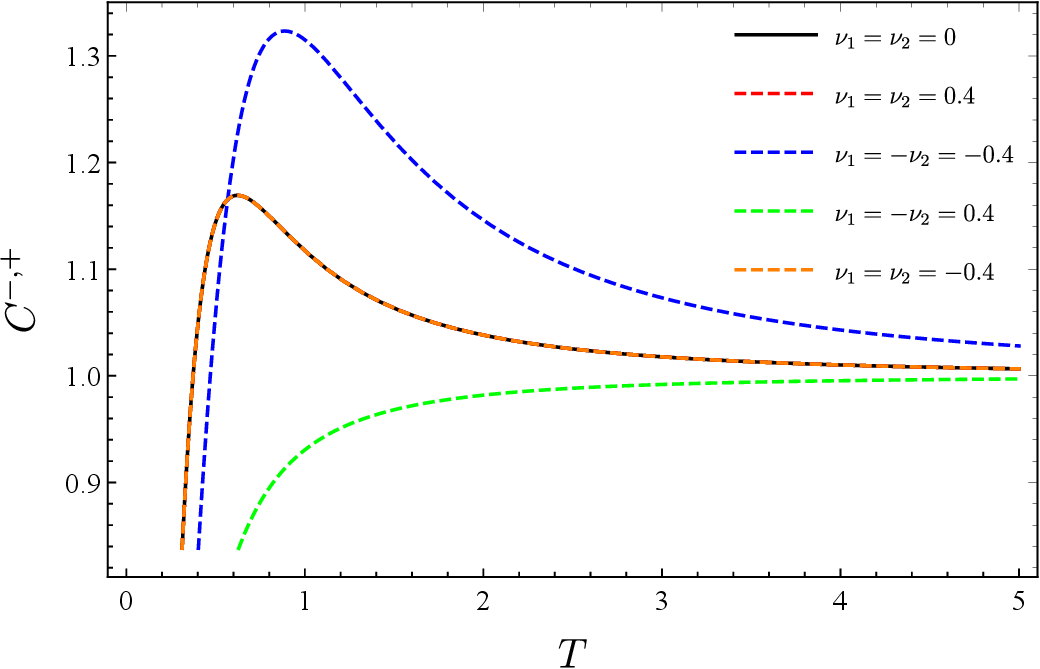}
       \subcaption{ $\epsilon _{1}=-1$, $\epsilon
_{2}=+1$ case.}
    \end{minipage}\hfill
\caption{A qualitative representation of the Dunkl-heat capacity versus temperature for different values of $%
\protect\nu _{1}$ and $\protect\nu _{2}$.} \label{Fig5}
\end{figure}

\begin{figure}[htb]
\begin{minipage}[t]{0.5\textwidth}
        \centering
        \includegraphics[width=\textwidth]{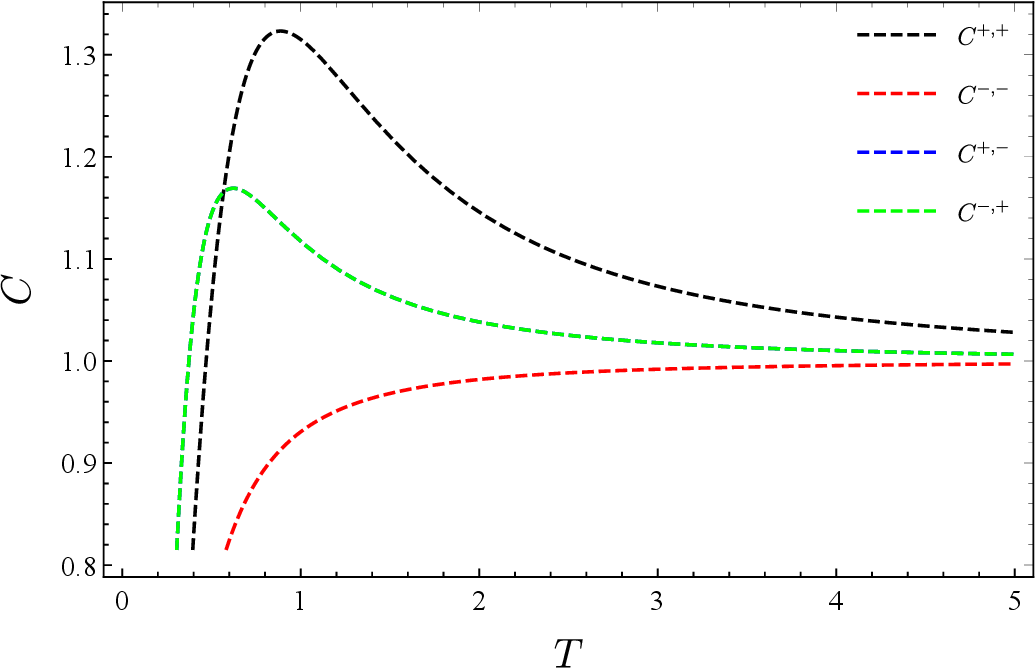}
       \subcaption{ $ \nu_{1}=0.4$, and $\nu_{2}=0.4$.}\label{fig:Ha}
   \end{minipage}%
\begin{minipage}[t]{0.50\textwidth}
        \centering
       \includegraphics[width=\textwidth]{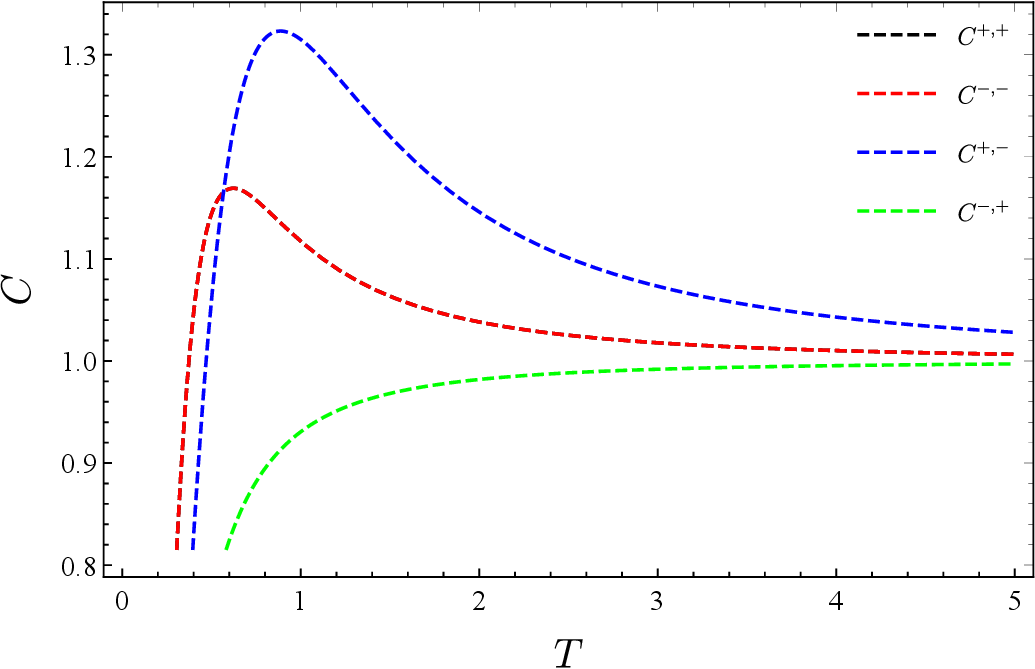}\\
        \subcaption{  $ \nu_{1}=0.4$, and $\nu_{2}=-0.4$.}\label{fig:Hb}
    \end{minipage}\hfill 
\begin{minipage}[b]{0.5\textwidth}
        \centering
        \includegraphics[width=\textwidth]{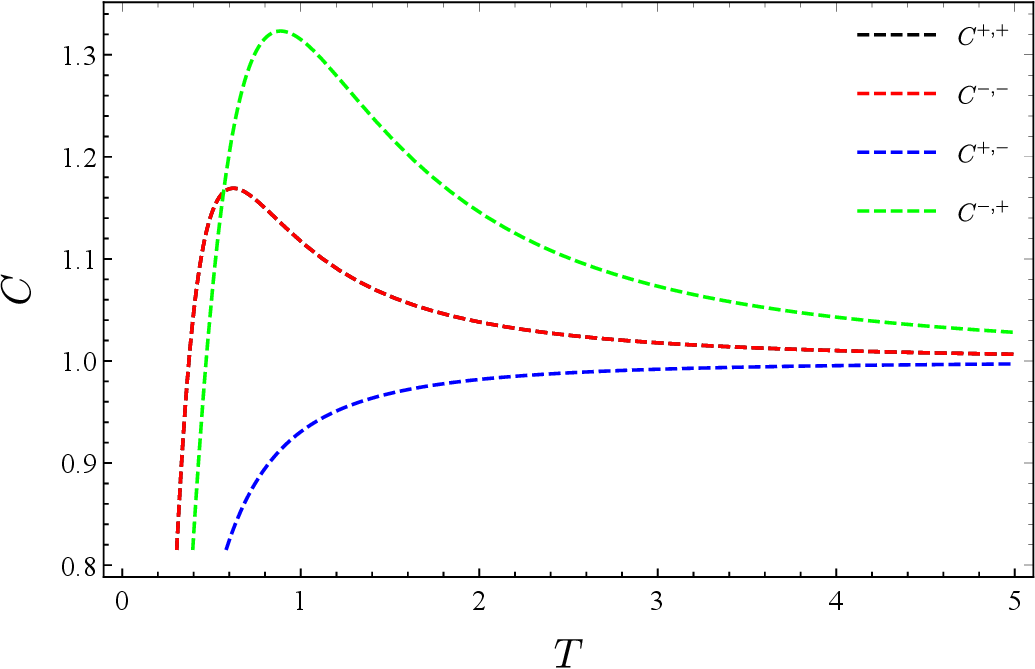}
       \subcaption{ $ \nu_{1}=-0.4$, and $\nu_{2}=0.4$.}\label{fig:Cj}
   \end{minipage}%
\begin{minipage}[b]{0.50\textwidth}
        \centering
       \includegraphics[width=\textwidth]{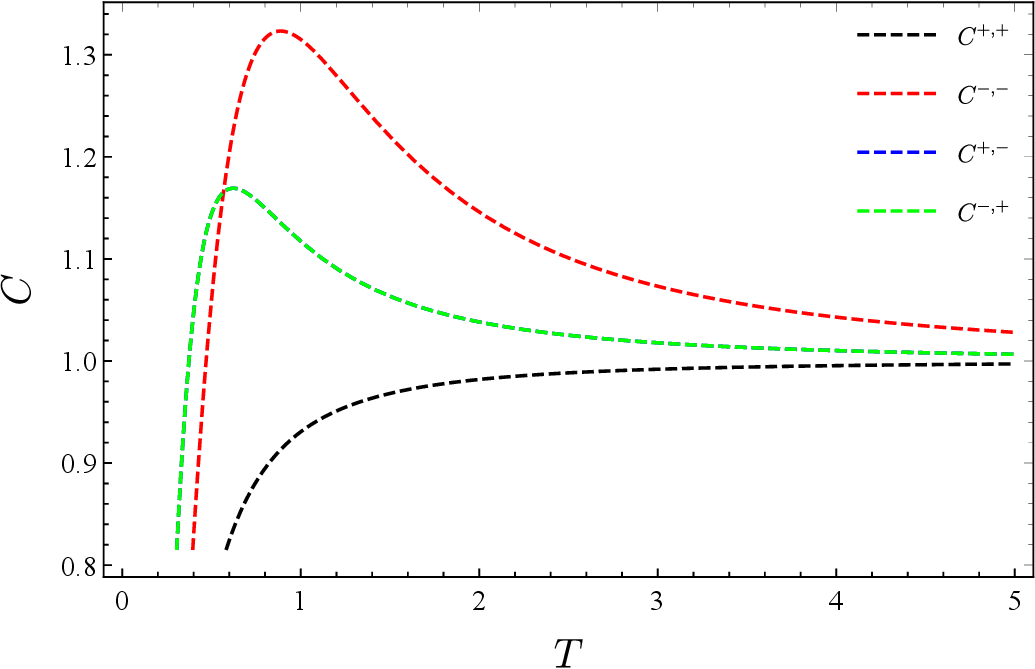}\\
        \subcaption{ $ \nu_{1}=-0.4$, and $\nu_{2}=-0.4$.}\label{fig:Cd}
    \end{minipage}\hfill
\caption{A qualitative representation of the Dunkl-heat capacity function versus temperature for constant Wigner parameters. } \label{Fig6}
\end{figure}

\newpage
Next, we calculate the Dunk-heat capacity function by using 
\begin{equation}
\frac{C}{K_{B}}=-\beta ^{2}\frac{\partial }{\partial \beta }U.  \label{CD}
\end{equation}%
After substituting Eq. (\ref{DU}) into Eq.(\ref{CD}) , we get the Dunkl-heat capacity in the form of
\begin{equation}
\frac{C_{\ell }^{\epsilon _{1},\epsilon _{2}}}{K_{B}}=\frac{\beta ^{2}\omega
_{c}^{2}}{4\sinh ^{2}\left( \frac{\beta \omega _{c}}{2}\right) }+\frac{%
\left( \beta \omega _{c}\eta ^{\epsilon _{1},\epsilon _{2}}\right) ^{2}}{%
\cosh ^{2}\left( \beta \omega _{c}\eta ^{\epsilon _{1},\epsilon _{2}}\right) 
}. \label{DHC}
\end{equation}%
In the high and low-temperature regimes, this formula reduces to
\begin{equation}
\frac{C_{\ell }^{\epsilon _{1},\epsilon _{2}}}{K_{B}}\simeq \left\{ 
\begin{array}{ll}
1,  & KT>>\omega _{c} \\ 
\left( \frac{\beta \omega _{c}}{2}\right) ^{2}e^{-\beta \omega _{c}}+\left(
\beta \omega _{c}\eta ^{\epsilon _{1},\epsilon _{2}}\right) ^{2}e^{-2\beta
\omega _{c}\eta ^{\epsilon _{1},\epsilon _{2}}}, & KT<<\omega
_{c}%
\end{array}%
\right. 
\end{equation}%
In Figs. \ref{Fig5} and \ref{Fig6},  we present how the Dunkl-heat capacity function changes with temperature for different Wigner parameter values.

\newpage
{ We observe} that as the temperature {escalates} the Dunkl-heat capacity increases until it reaches its maximum value, after which it decreases beyond a critical temperature value, $T_{c}$, in all cases. This indicates the possibility of phase transitions within the system. Additionally, each curve exhibits a distinct peak in heat capacity function as temperature crosses the critical temperature. We note that the maximum value of Dunkl's heat capacity and the critical temperature value alter with variations of the parameters $\nu_{j}$ and $\epsilon_{j}$. Moreover, we observe that at the high-temperature regime, all curves coincide, and they reach a fixed value,  $C=1$. 

\begin{figure}[htb]
\begin{minipage}[t]{0.5\textwidth}
        \centering
        \includegraphics[width=\textwidth]{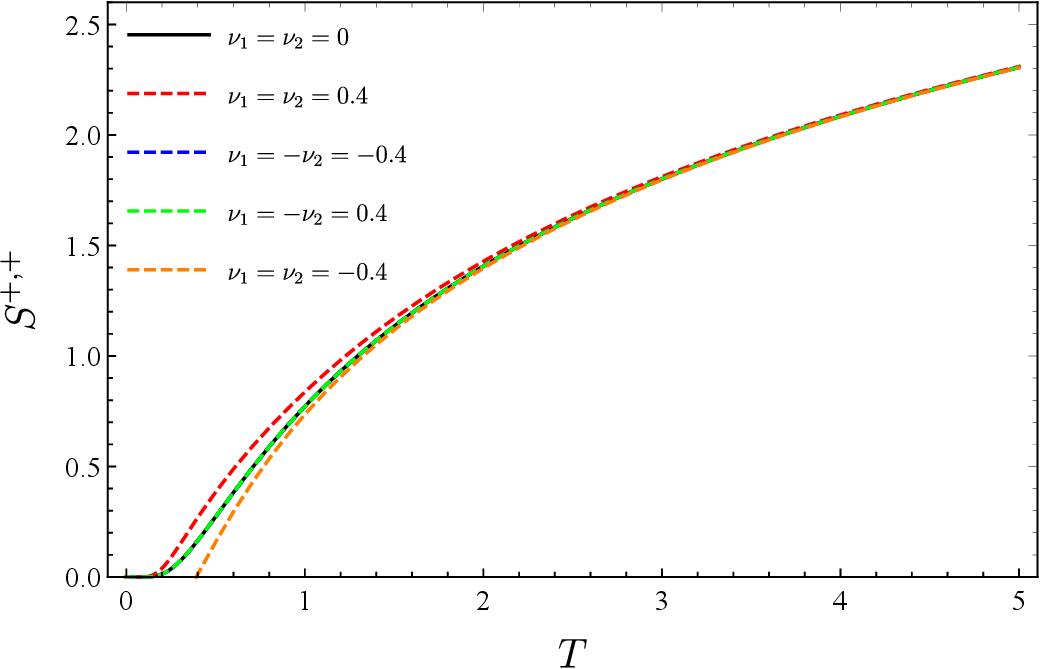}
         \subcaption{$\epsilon _{1}=\epsilon
_{2}=+1$ case. }
           \end{minipage}%
\begin{minipage}[t]{0.50\textwidth}
        \centering
       \includegraphics[width=\textwidth]{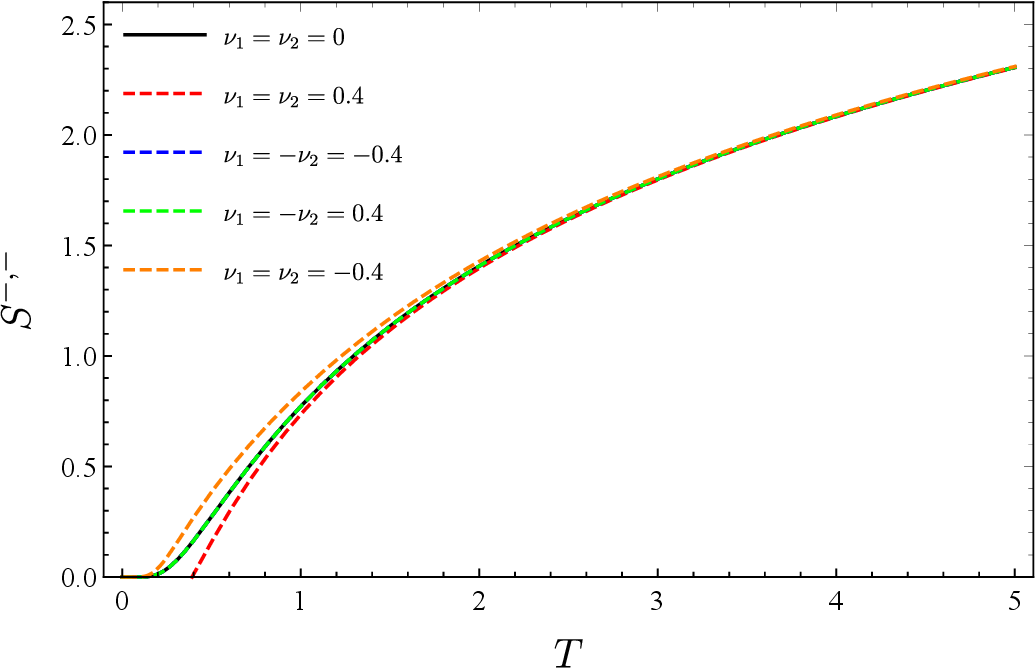}
        \subcaption{$\epsilon _{1}=\epsilon
_{2}=-1$ case. }
    \end{minipage}\hfill 
\begin{minipage}[t]{0.5\textwidth}
        \centering
        \includegraphics[width=\textwidth]{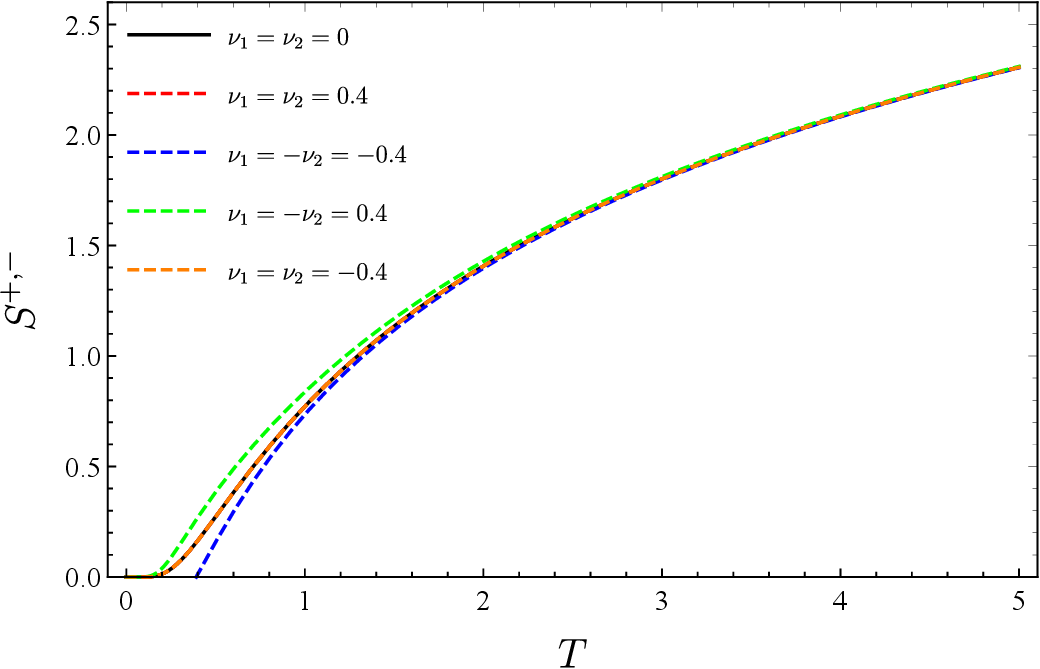}
\subcaption{ $\epsilon _{1}=+1$, $\epsilon
_{2}=-1$ case.}
   \end{minipage}%
\begin{minipage}[t]{0.50\textwidth}
        \centering
       \includegraphics[width=\textwidth]{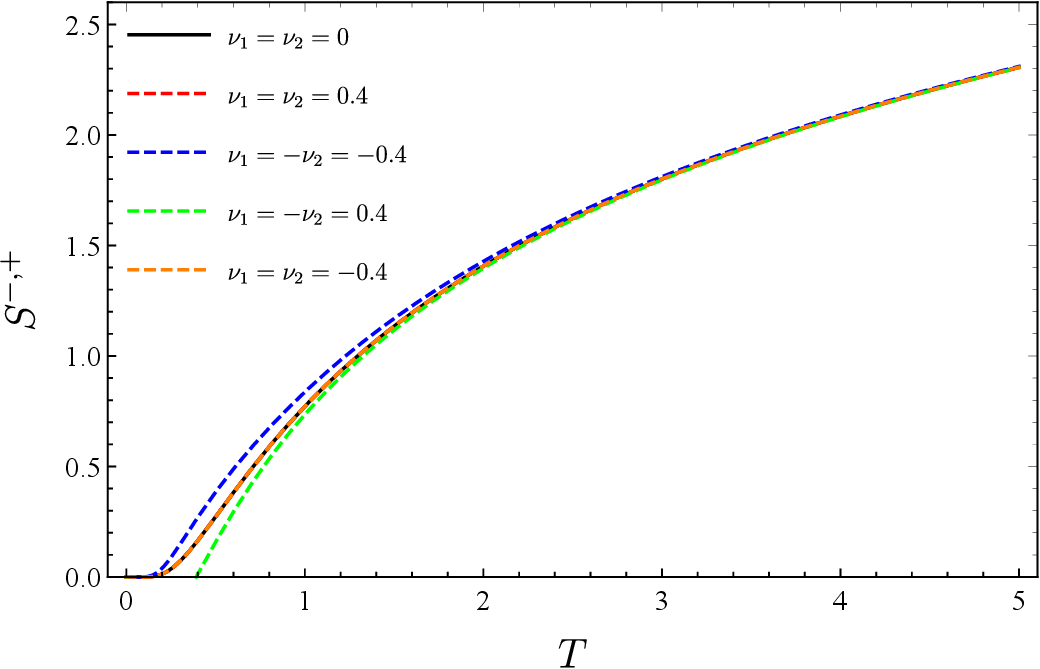}
\subcaption{ $\epsilon _{1}=-1$, $\epsilon
_{2}=+1$ case.}
    \end{minipage}\hfill
\caption{A qualitative representation of the Dunkl-entropy versus temperature for different values of $\protect%
\nu _{1}$ and $\protect\nu _{2}$.} \label{Fig7}
\end{figure}

\begin{figure}[htb]
\begin{minipage}[t]{0.5\textwidth}
        \centering
        \includegraphics[width=\textwidth]{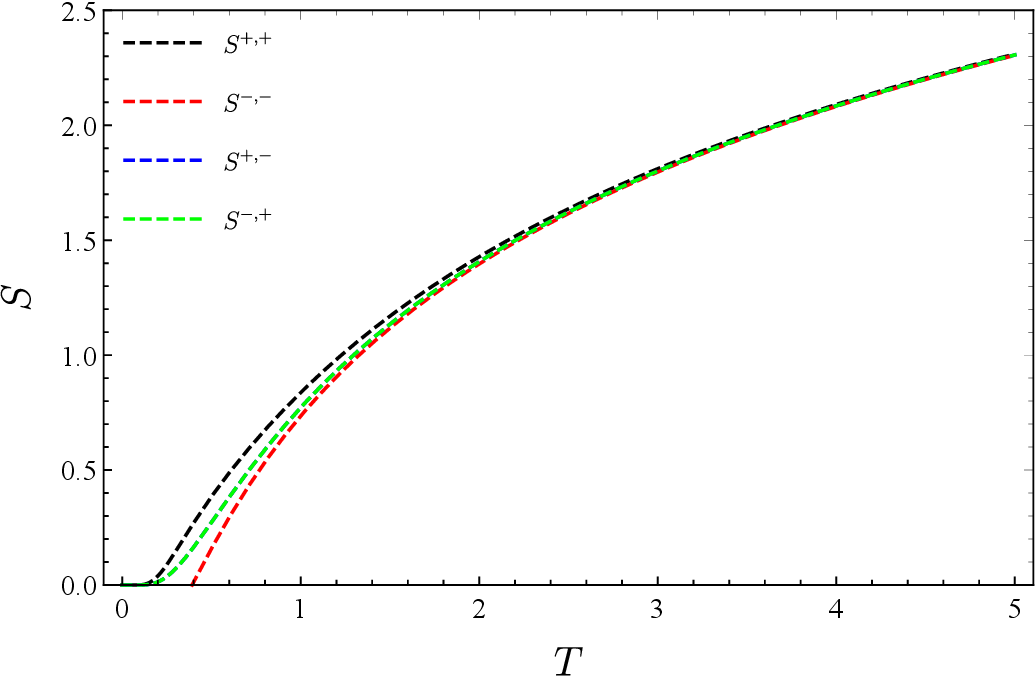}
       \subcaption{ $ \nu_{1}=0.4$, and $\nu_{2}=0.4$.}\label{fig:Sa}
   \end{minipage}%
\begin{minipage}[t]{0.50\textwidth}
        \centering
       \includegraphics[width=\textwidth]{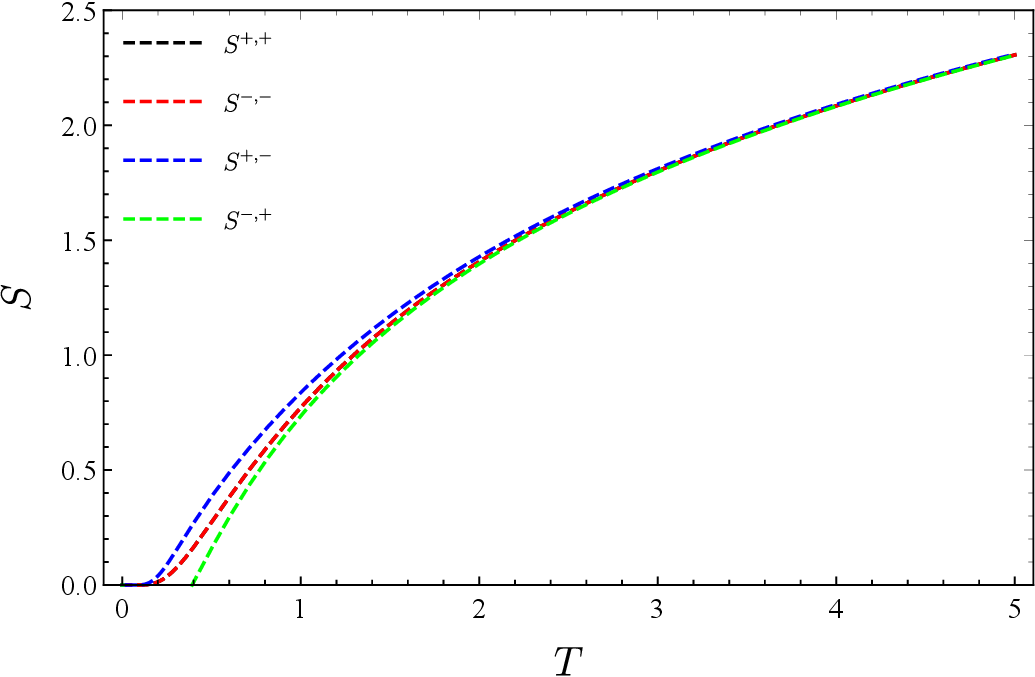}\\
        \subcaption{  $ \nu_{1}=0.4$, and $\nu_{2}=-0.4$.}\label{fig:Sb}
    \end{minipage}\hfill 
\begin{minipage}[b]{0.5\textwidth}
        \centering
        \includegraphics[width=\textwidth]{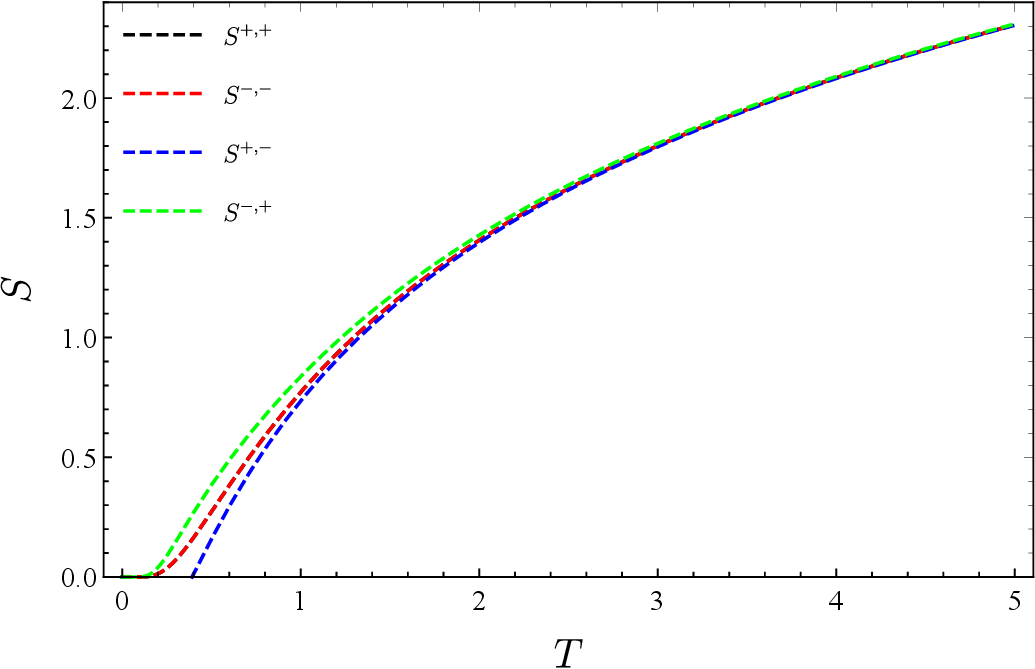}
       \subcaption{ $ \nu_{1}=-0.4$, and $\nu_{2}=0.4$.}\label{fig:Sc}
   \end{minipage}%
\begin{minipage}[b]{0.50\textwidth}
        \centering
       \includegraphics[width=\textwidth]{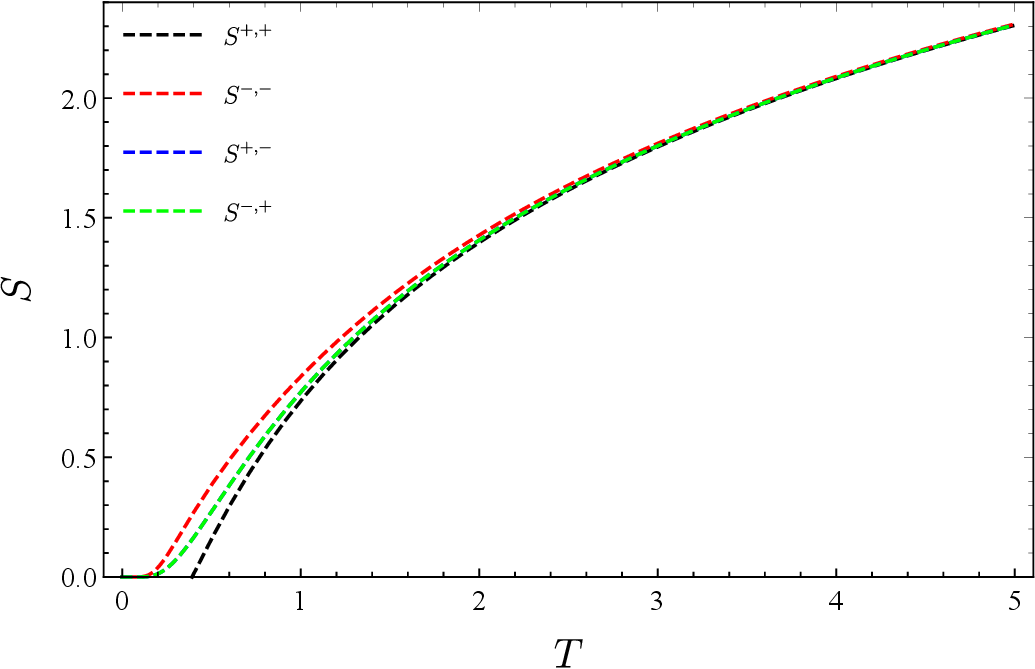}\\
        \subcaption{ $ \nu_{1}=-0.4$, and $\nu_{2}=-0.4$.}\label{fig:Sd}
    \end{minipage}\hfill
\caption{A qualitative representation of the Dunkl-entropy function versus temperature for constant Wigner parameters. } \label{Fig8}
\end{figure}

\newpage
Finally, we examine the reduced Dunkl-entropy {function} according to 
\begin{eqnarray}
\frac{S}{K_{B}}=\beta ^{2}\frac{\partial }{\partial \beta }F.    
\end{eqnarray}
{Here, we} obtain
\begin{equation}
\frac{S_{\ell }^{\epsilon _{1},\epsilon _{2}}}{K_{B}}=-\log \sinh \left( 
\frac{\beta \omega _{c}}{2}\right) +\frac{\beta \omega _{c}}{2}\coth \left( 
\frac{\beta \omega _{c}}{2}\right) +\log \cosh \left( \beta \omega _{c}\eta
^{\epsilon _{1},\epsilon _{2}}\right) -\left( \beta \omega _{c}\eta
^{\epsilon _{1},\epsilon _{2}}\right) \coth \left( \beta \omega _{c}\eta
^{\epsilon _{1},\epsilon _{2}}\right) , \label{DS}
\end{equation}
{and in} Figs. \ref{Fig7} and \ref{Fig8}, we {display} the Dunkl-entropy function versus temperature for different Wigner parameter values.

\newpage
We see that at low-temperature limits the Dunkl-entropy behaves differently for various values of $\epsilon_{j}$. However,  at high temperatures, there is no difference between the standard and Dunkl entropy. 

\section{Conclusion}
Similar to the Pauli equation, the Dunkl derivative, hence, the Dunkl formalism has a long historical background. However, in the last decade, Dunkl formalism has been used in various fields of physics since it allows us to obtain parity-dependent solutions. In this manuscript, we considered a nonrelativistic system of spin-$1/2$ particles in two dimensions, and in the presence of an external magnetic field, we looked for an analytical solution out of the Dunkl-Pauli equation. After the straightforward calculations, we obtained the eigenvalue and the corresponding eigenfunctions in terms of confluent hypergeometric functions. Then, we extended our work through the statistical mechanical framework. We defined the partition function by considering the system to be in thermal equilibrium. Then, we obtained the Dunkl formalism-modified Helmholtz free energy, internal energy, specific heat, and entropy functions, respectively. With the graphical discussions, we intended to present the effect of the latter formalism on the thermal quantities through different parity symmetries.

{ As part of future research directions, we suggest exploring the thermodynamic characteristics of Dunkl-fermions within the grand canonical ensemble framework. Here,  we propose using a methodology similar to recent literature \cite{rn1,rn2,rn3} for examining the phase transition mechanism and determining whether entropy and average energy are included in the Dunkl formalism. We have to emphasize that these ideas, along with others, are currently under development.}

\newpage
\section*{Acknowledgments}

{The authors are thankful to the anonymous reviewer for the constructive comments.} This work is supported by the Ministry of Higher Education and Scientific Research, Algeria under the code: B00L02UN040120230003. B C L\"{u}tf\"{u}o\u{g}lu is grateful to the P\v{r}F UHK Excellence project of 2211/2023-2024
for the financial support.

\section*{Data Availability Statements}

The authors declare that the data supporting the findings of this study are
available within the article.

\end{document}